\documentclass{article}

\usepackage{arxiv}

\usepackage[utf8]{inputenc} % allow utf-8 input
\usepackage[T1]{fontenc}    % use 8-bit T1 fonts
\usepackage{hyperref}       % hyperlinks
\usepackage{url}            % simple URL typesetting
\usepackage{booktabs}       % professional-quality tables
\usepackage{amsfonts}       % blackboard math symbols
\usepackage{nicefrac}       % compact symbols for 1/2, etc.
\usepackage{microtype}      % microtypography
\usepackage{amsmath}
\usepackage{subcaption}
\usepackage{graphicx}
\usepackage{natbib}
\usepackage{doi}

\title{Computational study of airfoil stall flutter Limit Cycle Oscillations}

\date{} 					% Or removing it

\author{ \href{https://orcid.org/0000-0001-9236-8133}           {\includegraphics[scale=0.06]{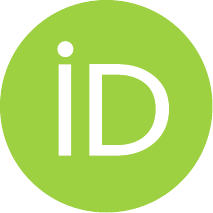}\hspace{1mm}Nikos Spyropoulos} \\
	School of Mechanical Engineering \\
	National Technical University of Athens \\
	Zografos, Athens, 15710, Greece \\
	\texttt{nspyro@fluid.mech.ntua.gr} \\
	\And
	\href{https://orcid.org/0000-0002-5506-6061}{\includegraphics[scale=0.06]{orcid.pdf}\hspace{1mm}Marinos Manolesos} \\
	School of Mechanical Engineering \\
    National Technical University of Athens \\
	Zografos, Athens, 15710, Greece \\
	\texttt{marinos@fluid.mech.ntua.gr} \\
	\AND
    \href{https://orcid.org/0000-0002-2742-5258}
    {\includegraphics[scale=0.06]{orcid.pdf}\hspace{1mm}George Papadakis} \\
	School of Naval Architecture and Marine Engineering \\
    National Technical University of Athens\\
	Zografos, Athens, 15710, Greece \\
	\texttt{papis@fluid.mech.ntua.gr} \\
}

\renewcommand{\shorttitle}{Prediction of Airfoil Stall Flutter}

\hypersetup{
pdftitle={A template for the arxiv style},
pdfsubject={q-bio.NC, q-bio.QM},
pdfauthor={David S.~Hippocampus, Elias D.~Striatum},
pdfkeywords={First keyword, Second keyword, More},
}

\begin{document}
\maketitle

\begin{abstract}
This paper presents a comprehensive numerical investigation of a NACA0012 undergoing Stall Flutter Limit Cycle Oscillations (LCO) across distinct fluid dynamics regimes. It accurately models Small Amplitude Oscillations (SAO) in the transitional Reynolds regime and Large Amplitude Oscillations (LAO) in the moderate regime, observed in different experimental campaigns. The SAO analysis serves as a verification of the computational framework against established numerical benchmarks. Crucially, the LAO simulations represent the first documented prediction across the full experimental velocity range correlated against available measured data, addressing a significant literature gap. The predictions fidelity relies on rigorous computational criteria defined through a detailed sensitivity analysis. This demonstrated numerical requirements significantly more demanding than those typically employed for computing static polars or simulating dynamic pitching motion of rigid airfoils, underscoring the severity of the aeroelastic problem. Quantitatively the simulation systematically over--predicts the critical onset velocity and under--predicts the LCO amplitudes.However, the results show strong qualitative agreement with experimental observations, successfully reproducing key dynamic stall mechanics and bifurcation phenomena.
\end{abstract}

% keywords can be removed

\section{Introduction}
\label{sec:introduction}
\begin{figure}
    \centering
    \includegraphics[width=0.75\linewidth]{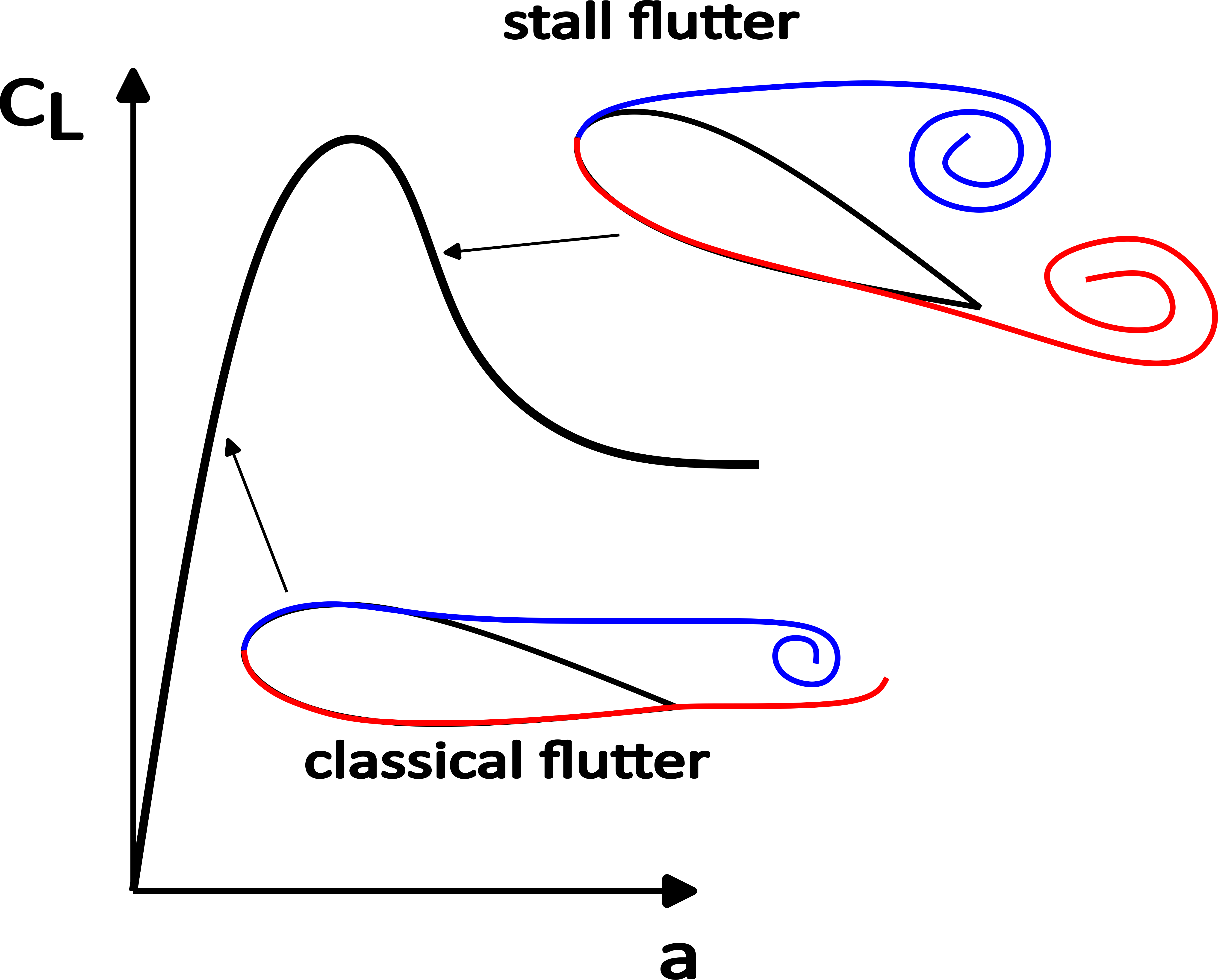}
    \caption{Schematic diagram of different flutter types.}
    \label{fig:flutter_types}
\end{figure}

Stall flutter is a non--linear aeroelastic instability phenomenon resulting from the coupling between the flexibility of an aerodynamic structure (a wing) and the flow, in cases fundamentally driven by dynamic flow separation (dynamic stall). The resulting structural response ends up in a change in the pitch angle of the wing, which in turn may give rise to periodic complex flow separation and perhaps re--attachment on both sides of the lifting surface. This interaction between the fluid and the structure (Fluid Structure Interaction -- FSI) creates self--excited oscillations and allows for the continued energy transfer from the free stream to the wing and vice--versa. Depending on the correlation between the aerodynamic energy input and the structural energy dissipation, the system’s dynamic behavior varies. Classical bending--torsion flutter involves the interaction between two (or more) structural modes, leading to self--excited vibrations of exponentially increasing amplitude that are mainly unrelated to flow separation (at least in the initial phase of the phenomenon) and usually occur in a linear flow regime (see Figure \ref{fig:flutter_types}). On the other hand, stall flutter is strongly related to non--linear aerodynamic excitation phenomena arising in stall and post--stall angles of attack, leading in many cases to self--excited single degree of freedom (pitching) oscillations of limited amplitude that are usually referred to as Limit Cycle Oscillations (LCO). LCO represent a state of dynamic equilibrium where the time--averaged energy absorbed from the fluid over one cycle is balanced by the energy dissipated by structural damping. This phenomenon is typically a single degree of freedom instability, that only occurs if the pitching eigen--frequency is lower than the plunging one, or if plunging is restricted. It is commonly encountered on similar aerodynamic shapes such as wings or rotating blades in many different applications, including helicopter rotor blades \cite{ham1966torsional}, wind turbine blades \cite{ericsson2000role}, turbomachinery blades \cite{mendelson1949aerodynamic}, aircraft tails \cite{victory1943flutter} and highly flexible wings \cite{noll2004investigation}.

Although initial experimental work on stall flutter dates back to the 1940s \cite{victory1943flutter}, much of the related research has historically focused on the aerodynamic phenomenon of dynamic stall in the second half of the $20^{th}$ century \cite{halfman1951evaluation,ericsson1980dynamic,mccroskey1981phenomenon,rivera1992naca}. Nevertheless, stall flutter is an aeroelastic phenomenon resulting in LCO introduced by complex and non--linear dynamic flow phenomena such as dynamic stall. Dynamic stall is a distinct unsteady aerodynamic phenomenon, in which an abrupt loss of aerodynamic loads occurs due to flow separation over a lifting surface undergoing unsteady motion that may not necessarily originate in aeroelastic responses. Yet, in stall flutter oscillations, dynamic stall serves as the main excitation factor that stabilizes the limited amplitude oscillations. Over recent years, various types of low--speed stall flutter oscillations have been observed experimentally, leading to several classifications, typically based on the dominant degree of freedom (e.g. pitch-dominated, plunge-dominated). Within the pitch-dominated classification, distinct behaviors are observed:
\begin{itemize}
    \item Small Amplitude Oscillations (SAO)
    \item Large Amplitude Oscillations (LAO)
\end{itemize}

SAO with symmetric pitch response in the range between $0^{\circ}$ and $10^{\circ}$ around $0^{\circ}$ mean value were first systematically observed and reported through wind tunnel experiments by \cite{poirel2008self} in the transitional Reynolds number regime $(4.5*10^4 \leq Re_c \leq 1.3*10^5)$. The loss of the dynamic stability at $0^{\circ}$ pitch angle was initially attributed to the laminar boundary separation near the trailing edge region, while the subsequent presence of a Laminar Separation Bubble (LSB) was considered as a stabilisation factor of the fluttering motion into LCO. The presence of a LSB is computationally verified in \cite{poirel2010aerodynamics} and in the present study (see Section \ref{ssec:low_amplitude}). Nevertheless, simulations indicate a Kelvin--Helmholtz instability occurring in the shear layer close to the trailing edge as the onset mechanism to the fluttering motion and the resulting von Kármán vortex shedding as the LCO stabilisation factor. This transitional instability regime has since been thoroughly investigated using various Computational Fluid Dynamics (CFD) tools. Early computational efforts to numerically investigate the flow patterns generated around the oscillating wing and the resulting aerodynamics loads utilised Large Eddy Simulations (LES) relying on imposed motion \cite{poirel2010aerodynamics}, while later analyses successfully employed fully unsteady aeroelastic simulations based on a 2D Unsteady Reynolds Averaged Navier Stokes (URANS) approach, effectively reproducing the self--sustained pitch oscillations across the relevant Reynolds number range \cite{poirel2011computational,abbas2021aeroelastic}.  These numerical predictions are utilised in the present work for the verification of our comprehensive computational framework, allowing for comparison against external data of comparable fidelity.

LAO with symmetric pitch response in the range between $10^{\circ}$ and $40^{\circ}$ were later observed by \cite{li2007experimental,dimitriadis2009bifurcation} in a higher Reynolds number range $(2.5*10^5 \leq Re_c \leq 5*10^5)$ and were linked to the generation and shedding of a strong Leading Edge Vortex (LEV), similar to those observed during dynamic stall. In the same experimental campaign, asymmetric LCO of smaller amplitude (up to $20^{\circ}$) around the static equilibrium angle have been also observed at higher Reynolds numbers (up to $5*10^5 \leq Re_c \leq 6*10^5$), resulting from the interaction between the stall flutter mechanism and the system’s static divergence. Contrary to SAO cases, accurate CFD predictions reproducing these large amplitude LCO experiments are scarce in the literature. At least to the authors' knowledge, the successful computational replication of these large amplitude stall flutter LCO across the entire experimental Reynolds number range has not yet been thoroughly reported, highlighting a significant literature gap that this paper aims to address.

This paper presents a comprehensive numerical investigation focused on accurately capturing both small and large amplitude LCO arising from Stall Flutter instabilities in both low and moderate Reynolds number ranges. The computational framework presented is initially verified using the SAO experiments \cite{poirel2008self}. In addition to wind tunnel measurements, present study results are compared to published numerical predictions of similar fidelity tools available  in the literature. Subsequently, the approach extends to large amplitude LCO in the higher Reynolds number range \cite{dimitriadis2009bifurcation}, which is closer to actual aeronautical applications. Accurate CFD predictions reproducing these specific large amplitude LCO have not yet been reported to literature for the full Reynolds number range of the experimental campaign. The aeroelastic responses of both experimental campaigns are computed using the in--house solver MaPFlow \cite{Papadakis_phd}, which solves the compressible URANS equations combined with Low Mach Preconditioning for robust and accurate FSI computations in the low Mach number flow regions. For the successful simulation of the SAO, which are inherently linked to the laminar separation flutter mechanism, the URANS methodology explicitly incorporates the $\gamma$--$Re_\theta$ transition model of Menter \cite{langtry2009correlation}, whereas fully turbulent simulations using the (k--$\omega$ SST) two equations turbulence model of Menter \cite{menter1994two} have been conducted in LAO cases. Furthermore, a detailed sensitivity analysis of the computational methodology is undertaken, specifically focusing on the investigation of the most significant numerical parameters (grid and time--step resolution) under the demanding conditions of large amplitude aeroelastic oscillations. This rigorous analysis demonstrates the necessity of investigating parameters such as minimum spacing at the leading and trailing edges to achieve grid--independent LCO predictions, a process often overlooked when analysis is restricted solely to computing static polars \cite{watrin2012computational,yabili2012unsteady,poirel2011computational,abbas2021aeroelastic,li2024mechanism}. Furthermore, it establishes rigorous temporal requirements significantly stricter than those commonly found sufficient for dynamically pitching rigid airfoils.

The structure of the paper is as follows. Section \ref{sec:exp_setup} describes the experimental setups used for validation, focusing on small and large amplitude oscillations. Section \ref{sec:methodology} details the computational methodology, including the aerodynamic and structural models used, the aeroelastic coupling approach, and the comprehensive sensitivity analysis performed for grid and time--step dependency. Section \ref{sec:results} presents the numerical results, comparing predicted LCO characteristics and flow physics against experimental measurements and computational predictions (wherever available) for both small and large amplitude stall flutter cases. Finally, Section \ref{sec:conclusions} concludes the paper by summarizing the key findings and contributions.

\section{Experimental set--up}
\label{sec:exp_setup}
The following sections detail the experimental configurations utilised for the validation the computational framework presented in this paper, focusing on the distinct stall flutter regimes: Small Amplitude Oscillations (SAO) and Large Amplitude Oscillations (LAO). Sections \label{subsec:LAO_exp_setup} and \ref{subsec:LAO_exp_setup} provide specific details on the physical set--ups and structural parameters of these experiments, while Section \ref{subsec:comp_setup}, offers a general overview of the numerical approach.

\subsection{Small Amplitude Oscillations}
\label{subsec:SAO_exp_setup}
\begin{figure}
    \centering
    \includegraphics[width=0.75\linewidth]{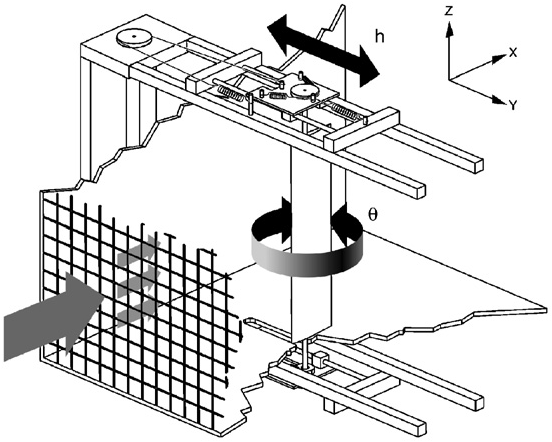}
    \caption{Schematic of the experimental set--up used to study Small Amplitude Stall Flutter Oscillations. Image adapted from \cite{poirel2008self}.}
    \label{fig:poirel_experimental_setup}
\end{figure}

The validation case for SAO is based on experiments performed by Poirel, Harris and Benaissa \cite{poirel2008self}. This study focused on aeroelastic LCO occurring in the narrow transitional Reynolds number regime, specifically $4.5*10^4 \leq Re_c \leq 1.3*10^5$. The model utilised a NACA0012 airfoil section (chord $c=0.156 \; m$, span $b=0.61 \; m$) that was structurally constrained to rotate in pure pitch, with plunging motion held fixed due to infinite stiffness, and was mounted with end--plates in order to minimize $3D$ effects. The apparatus possessed a mass moment of inertia $I_s=0.00135 \; kgm^2$ and a structural stiffness $K_s=0.30 \; Nm/rad$ and in the baseline configuration, the elastic axis (pitching axis) was located at $X_{ea}=0.18c$ aft of the leading edge. The system exhibited a viscous structural damping coefficient $D_s = 0.002 \; Nms$, resulting in a structural damping ratio $\zeta = 0.05$ during free decay tests in still air. The self--sustained oscillations were typically observed for airspeeds between $5$ and $12 \; m/s$, corresponding to the transitional Reynolds range. The baseline value for the free--stream turbulence intensity of the wind tunnel was $T_u \approx 0.2\%$. These small amplitude oscillations ($\theta _{max} \leq 5.5^{\circ}$) were the first LCO associated with low Reynolds number effects to be systematically reported. The underlying mechanism is fundamentally linked to laminar boundary layer separation near the trailing edge, initiating the instability and leading to the designation laminar separation flutter. The motion was generally characterized by a well--behaved harmonic response and was measured with rotary potentiometers, whereas the flow periodicity was monitored using hot--wire anemometry in the wake. The aerodynamic moment was calculated as the right hand side of the dynamic equation of motion of the wing, by assuming a pure sinusoidal variation of the pitch angle of the wing with a known frequency and amplitude. A schematic of the experimental set--up is shown in Figure \ref{fig:poirel_experimental_setup}.

\subsection{Large Amplitude Oscillations}
\label{subsec:LAO_exp_setup}
\begin{figure}
    \centering
    \includegraphics[width=0.75\linewidth]{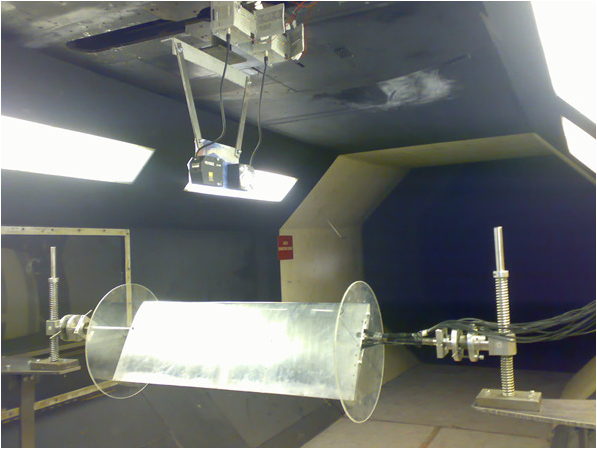}
    \caption{Image of the experimental set--up used to study Large Amplitude Stall Flutter Oscillations. Image copied from \cite{li2007experimental}.}
    \label{fig:dimitriadis_experimental_setup}
\end{figure}

The LAO experiments used for validation were conducted at the University of Manchester by Dimitriadis and Li \cite{li2007experimental,dimitriadis2009bifurcation}. Static measurements were carried out in the Project tunnel, which is an open--return subsonic wind tunnel with a working section of $2.16 \; m$ in length, $1.1 \; m$ in width and $0.86 \; m$ in height, whereas dynamic measurements were conducted in the Avro $9 x 7 \; ft^2$ closed--return subsonic wind tunnel. The test model featured a rectangular wing with a NACA0012 airfoil section, defined by a $b=900 \; mm$ span and a $c=300 \; mm$ chord. The support mechanism allowed for two degrees of freedom (pitch and plunge). However, the configuration was engineered to exhibit a single--mode dynamic response dominated by pitch motion due to the intentionally high plunge stiffness ($K_h=30.5 \; N/mm$) compared to the pitch stiffness ($K_p=13.1 \; Nm/rad$). The pitch axis was set $X_{ea} \approx 0.383c$ aft of the leading edge. The wing had a total weight of $m=15 \; kg$ and a moment of inertia of $I_p=0.31 \; kgm^2$ about the pitch axis. The system did not apply explicit damping mechanisms. Instead, the observed damping was primarily due to friction in the pitch springs, which led to non--linear structural damping. Analysis of free decay tests at $0$, $2.5$ and $3.1 \; m/s$ airspeed showed that the damping of the pitching mechanism was essentially linear at high pitch angles but became significantly non--linear at low pitch angles. The damping in plunge can be assumed to be very large. This campaign captured symmetric LCO with amplitudes reaching up to $40^{\circ}$, typically observed for airspeeds between $12$ and $25 \; m/s$ that correspond to a moderate Reynolds number range extending from $2.4*10^5$ up to $5*10^5$. Furthermore, asymmetric LCO were observed at higher airspeeds between $25$ and $30 \; m/s$ ($5*10^5 \leq Re_c \leq 6*10^5$) resulting from the interaction of the flutter mechanism with the static divergence airspeed of the system. No self--sustained oscillations were observed for airspeeds between $3$ and $12 \; m/s$ ($6*10^4 \leq Re_c \leq 2.4*10^5$). The free--stream turbulence intensity in this large wind tunnel was found to be low, typically estimated to be around $T_u=0.05\%$. Unsteady pressure data were measured around the mid--span section, whereas the motion of the wing was measured using laser displacement probes. A picture of the experimental set--up is shown in Figure \ref{fig:dimitriadis_experimental_setup}.

\subsection{Computational set--up}
\label{subsec:comp_setup}
\begin{figure}
    \centering
    \includegraphics[width=0.75\linewidth]{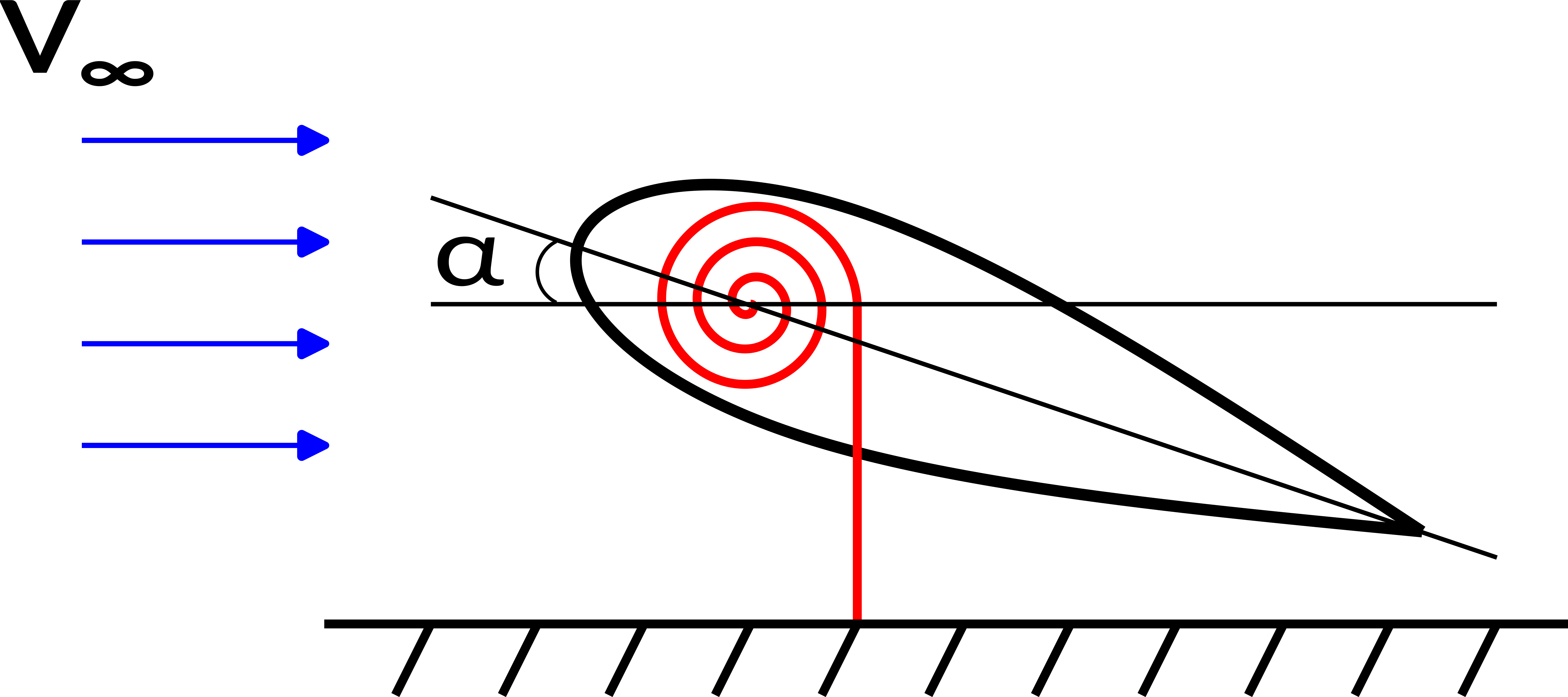}
    \caption{Schematic of the computational set--up used to study the Small Amplitude Stall Flutter Oscillations measured in \cite{poirel2008self} and the Large Amplitude Stall Flutter Oscillations measured in \cite{dimitriadis2009bifurcation}. The aerodynamic analysis is performed through a 2D URANS CFD framework, whereas for the prediction of the system dynamics response, only the pitch DOF is considered.}
    \label{fig:computational_setup}
\end{figure}

The numerical investigations detailed in the subsequent sections utilize the in--house CFD solver MaPFlow \cite{Papadakis_phd}. In this study, MaPFlow is preferred to solve the compressible URANS equations in a 2D context. This choice is supported by preceding computational studies which have successfully demonstrated the ability of RANS based aeroelastic solvers to capture many of the key features and flow physics associated with stall flutter \cite{poirel2011computational,yabili2012unsteady,abbas2021aeroelastic,li2024mechanism}. For robust and accurate FSI computations in the inherently low Mach number flow regions, Low Mach Preconditioning \cite{vatsa2004assessment} is consistently applied to the flow equations. The aerodynamic solver is strongly coupled with an internal $6$ Degrees Of Freedom (DOFs) Rigid Body Dynamics (RBD) solver. Nevertheless, in this investigation the computational model is restricted to the pure pitch (1 DOF) (see Figure \ref{fig:computational_setup}). This omission of the plunge DOF is justified, because experimental research confirmed that the plunging motion does not participate significantly in the stall flutter mechanism \cite{dimitriadis2009bifurcation}. Regarding turbulence modeling, the crucial distinction between the two regimes dictates the choice of models: fully turbulent simulations using the two equations turbulence model of Menter (k--$\omega$ SST) \cite{menter1994two} are conducted for the higher--Reynolds LAO; conversely, the $\gamma$--Re$\theta$ transitional model of Menter \cite{langtry2009correlation} is also employed for the low--Reynolds SAO, as this inclusion is necessary to capture the laminar separation flutter.

\section{Computational Methodology}
\label{sec:methodology}
This chapter presents the comprehensive computational methodology employed to simulate the aeroelastic response of the NACA0012 airfoil under stall flutter conditions, as defined by the experimental setups in Section \ref{sec:exp_setup}. The foundation of this analysis is the in--house CFD solver MaPFlow, which is strongly coupled to an internal 6 DOFs RBD solver for the analysis of FSI problems.

\subsection{The aerodynamic model}
\label{ssec:aero}

\newcommand{\partialt} [1] {\frac{\partial \emph{#1} }{\partial t}}
\newcommand{\vpartial} [2] {\frac{\partial\vec{ \emph{#1}} }{\partial \emph{#2}}}
\newcommand{\vpartialt}[1] {\frac{\partial\vec{ \emph{#1}} }{\partial t}}
\newcommand{\vfullt}   [2] {\frac{d\vec{ \emph{#1}_{\emph{#2}} }}{dt}}
\newcommand{\Vol}{D}

The core aerodynamic calculations for the aeroelastic simulations are performed using the in--house CFD solver MaPFlow \cite{papadakis2014development}, which utilises a cell--centered finite volume discretisation scheme and is capable of handling both structured and unstructured computational grids. MaPFlow is built upon the compressible URANS equations described by Equation (\ref{eq:URANS}). 
\begin{equation}
    \frac{\partial }{\partial t} \int_\Vol \mathbf{U} d\Vol + \oint_{\partial{\Vol}}\left(\mathbf{F_c}-\mathbf{F_v}\right)dS = \int_\Vol  \mathbf{Q}d\Vol 
\label{eq:URANS}
\end{equation}
In Equation (\ref{eq:URANS}), $\vec{U}$ is the vector of the Conservative Flow Variables: 
\begin{align*}
	\mathbf{U} = \left(\begin{array}{ccc}
	\rho   \\
	\rho u \\
	\rho v \\ 
	\rho w \\
	\rho E 
	\end{array} \right)
\end{align*}
where $\rho$ denotes density, $\left(u,v,w\right)$ the three components of the velocity field and $E$ the total energy.

The convective fluxes $\mathbf{F_c}$, described by Equation (\ref{eq:F_c}), are evaluated using the Roe’s approximate Riemann solver \cite{roe1986characteristic} described in Equation (\ref{eq:Roe}) coupled with the Venkatakrishnan limiter \cite{venkatakrishnan1993accuracy}.
\begin{equation}
	\mathbf{F_c}  = \left(\begin{array}{ccc}
	\rho V  \\
	\rho uV + n_xp \\
	\rho vV + n_yp \\ 
	\rho wV + n_zp \\
	\rho (E+\frac{p}{\rho}) V
	\end{array} \right)
\label{eq:F_c}
\end{equation}	

\begin{equation}
	\mathbf{F}_{c_f} = \frac{1}{2} [\mathbf F_c(\mathbf V_R)+\mathbf F_c(\mathbf V_L) - |\mathbf{A_{Roe}}|_f(\mathbf V_R - \mathbf V_L)]
\label{eq:Roe}
\end{equation}
The Left and Right states $\left(\mathbf V_L,  \mathbf V_R\right)$ of the primitive variables $\left(\mathbf{V}=\left(\rho,u,v,w,p\right)^T\right)$ of the face are computed using a Piecewise Linear Reconstruction (PLR) scheme and $|\mathbf{A_{Roe}}|$ is constructed using the absolute values of the eigenvalues and the right eigenvector matrix $R$:
\begin{equation*}
	|\mathbf{A_{Roe}}|=\mathbf R^{-1} |\pmb{\Lambda}| \mathbf R
\label{eq:roematshort}
\end{equation*}
In case Low Mach Preconditioning is used, the eigenvalues and the eigenvectors of the preconditioned system must be used ($\pmb{\Lambda}_\Gamma,\mathbf{R}_\Gamma $). Hence,  $|\mathbf{A_{Roe}}|$ is changed to (according  to \cite{Asouti_phd}):
\begin{align*}
	|\mathbf{A}_{\Gamma \mathbf{Roe}}| &=|\pmb{\Gamma}^{-1} \pmb{\Gamma} \mathbf{A_{Roe}}| \nonumber\\
	&   \simeq \pmb{\Gamma}^{-1} | \pmb{\Gamma} \mathbf{A_{Roe}}|\nonumber\\
	&   \simeq \pmb{\Gamma} ^{-1}  \mathbf{R}^{-1}_\Gamma |\pmb{\Lambda}_\Gamma| \mathbf{R}_\Gamma
\label{eq:roematshort2}
\end{align*}
	
The viscous fluxes $\mathbf{F_v}$, described by Equation (\ref{eq:F_v}), are calculated using a central $2nd$ order scheme.
\begin{equation}
	\mathbf{F_v} = \left(\begin{array}{ccc}
	0      \\
	n_x\tau_{xx}  + n_y\tau_{xy}  +n_z\tau_{xz}\\
	n_x\tau_{yx}  + n_y\tau_{yy}  +n_z\tau_{yz}\\
	n_x\tau_{zx}  + n_y\tau_{zy}  +n_z\tau_{zz}\\
	n_x\Theta_x + n_y\Theta_y +n_z\theta_z
	\end{array} \right)
\label{eq:F_v}
\end{equation}

In Equations (\ref{eq:F_c}) and (\ref{eq:F_v}) $V = \mathbf u \cdot \mathbf n$, $p$ denotes pressure, $\pmb{\tau}$ is the viscous stress tensor and 
\begin {align*}
	\Theta_x &= u\tau_{xx} + v\tau_{xy} + w\tau_{xz} \\
	\Theta_y &= u\tau_{yx} + v\tau_{yy} + w\tau_{yz} \\
	\Theta_z &= u\tau_{zx} + v\tau_{zy} + w\tau_{zz}
\end {align*}

The above system is completed with the Equation (\ref{eq:4}) of state for perfect gases: 
\begin{equation}
	p = (\gamma -1 )\rho \left[E - \frac{u^2 + v^2 + w^2}{2}\right]
\label{eq:4}
\end{equation}
	
Turbulence closures include the one equation turbulence model of Spalart--Allmaras (SA) \cite{spalart1992one}, as well as the two equations turbulence model of Menter ($k$--$\omega$ SST) \cite{menter1994two}. Regarding laminar--to--turbulent transition modelling, various models have been implemented, such as the correlation $\gamma$--$Re_ \theta$ model of Menter \cite{langtry2009correlation} and the $e^N$ model described in \cite{van1956suggested}. The Delayed Detached Eddy Simulation (DDES) approach implemented in MaPFlow, follows the suggestions of \cite{shur2008hybrid}.

MaPFlow can handle both steady and unsteady flows. The unsteady nature of FSI problems necessitates time--true computations using the implicit $2nd$ order backwards differences scheme described in Equation \eqref{eq:time_steady}, combined with the dual time--stepping technique to facilitate convergence \cite{vatsa2010re}.

\begin{equation}
	\frac{d\left(\Vol _I \mathbf U_I\right)}{dt} + \mathbf R^{n+1}_I = \mathbf{0}
\label{eq:time_steady}
\end{equation}
where
$$\frac{d\left(\Vol _I \mathbf U_I\right)}{dt} = \frac{1}{\Delta t} \left[ 3/2 \left(\Vol _I  \mathbf U _I \right)^{n+1} -2 \left(\Vol _I  \mathbf U _I \right)^n	+ 1/2 \left(\Vol _I \mathbf U _I \right)^{n-1} \right],$$
$$\mathbf R^{n+1} \approx \mathbf R^n + \left(\frac{\partial \mathbf R}{\partial \mathbf U} \right)_n \Delta \mathbf U^n,$$
$$\mathbf R_I= \left[ \sum\limits_{f=1}^{N_f} \left(\mathbf{F}_{c_f}-\mathbf{F}_{v_f}\right) \Delta S_f - D\mathbf{Q} \right]_I$$ and $$\Delta \mathbf U^n = \mathbf U^{n+1} -\mathbf U^{n}$$

To ensure numerical stability and robustness in the low Mach number flow regime, Eriksson's Preconditioning Matrix \cite{Eriksson1996} is consistently applied to the governing equations, due to its successful use in \cite{Asouti2005}.

\subsection{The structural dynamics model}
\label{ssec:struct}
The structural response of the airfoil is handled internally by an in--house RBD solver, which is integrated within the numerical framework MaPFlow and is strongly coupled with the aerodynamic system to perform high--fidelity FSI computations. For the validation study presented here, the RBD solver is strictly constrained to model only the pitch DOF. The governing equation of motion for pitch rotation $\theta$ is expressed as a standard second--order system:
\begin{equation}
    I\ddot{\theta}+C\dot{\theta}+K\theta=M_{aero}
\label{eq:RBD}
\end{equation}
, where $I$ is the mass moment of inertia, $C$ the damping coefficient and the $K$ is the stiffness of the torsional spring. $M_{aero}$ is the aerodynamic moment calculated by properly integrating pressure and viscous stresses over the boundary of the airfoil following equation \eqref{eq:mom}:
\begin{align}
   M_{aero} &= \int_{\partial B} \mathbf{r} \times ( -p \mathbf{n} + \pmb{\tau} \cdot \mathbf{n} ) dS
\label{eq:mom}
\end{align}
, where the $p$ is the pressure distribution over the lifting surface, $\pmb{\tau}$ is the viscous stress tensor, $\mathbf{n}$ the outwards pointing unit normal vector of the surface, $\mathbf{r}$ the position vector with its origin being at the center of rotation and $\partial B$ is the surface boundary of the rigid body.

The RBD solver integrates Equation (\ref{eq:RBD}) using the $2nd$ order implicit Newmark--$\beta$ method \cite{Bathe2006}.

\subsection{The aeroelastic coupling}
\label{ssec:coupling}
\begin{figure}
    \centering
    \includegraphics[width=0.5\linewidth]{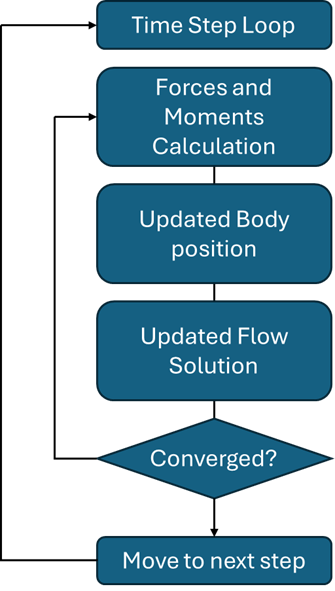}
    \caption{Flow chart of the fluid--structure coupling. Within every physical time step, internal iterations between the aerodynamics and rigid body dynamics solver are made to ensure a strong coupling and high--fidelity FSI computations.}
    \label{fig:aeroelastic_coupling}
\end{figure}

A strong aeroelastic coupling between the fluid domain and the structural dynamics solver is achieved through an iterative solution of the two separate, fluid and structural, problems within every physical time step to ensure convergence of the fluid--structure equilibrium. The overall FSI process involves the following exchange:
\begin{enumerate}
    \item Given a flow field solution, the CFD solver computes the aerodynamic moment ($M_{aero}$) acting on the airfoil.
    \item The aerodynamic moment is transferred to the RBD solver, which integrates the equation of motion (\ref{eq:RBD}) to determine the new pitch angle $\theta$, angular velocity $\dot{\theta}$ and angular acceleration $\ddot{\theta}$ of the airfoil.
    \item  The updated kinematic state (orientation and velocity) is communicated back to the flow solver by properly rotating the computational grid.
    \item The iterative loop continues until the calculated motion change with respect to the previous iteration achieves a predetermined tolerance of $\epsilon=10^{-4}$: $\max(|{\theta}^{n+1}-{\theta}^n|,|{\dot{\theta}}^{n+1}-{\dot{\theta}}^n|,|{\ddot{\theta}}^{n+1}-{\ddot{\theta}}^n|) \leq \epsilon$.
\end{enumerate}
The typical workflow of this iterative, strongly coupled procedure is detailed in Figure \ref{fig:aeroelastic_coupling}.
 
It needs to be highlighted that the computational mesh used in this study is rotated rigidly to accommodate the computed pitching motion. This approach is preferred to mesh deformation techniques as it conserves the regularity of the grid, thus reducing the numerical diffusion associated with mesh deformation, and ensures ease of implementation, thus restraining numerical errors. The rigid body rotation is implemented by employing the 2D rotation matrix $R(\theta)$ and its time derivative $\dot{R}(\theta,\dot{\theta})$ to calculate the new coordinates and velocities of the grid nodes.

\subsection{Sensitivity analysis}
\label{ssec:numerical}
In this section, a comprehensive sensitivity analysis of the employed numerical parameters is presented, aiming at deminishing numerical uncertainties and ensuring the robustness and validity of the predicted results. The core objectives of this section involve the rigorous investigation of grid and temporal discretisation effects. This study focuses specifically on a highly challenging LAO case at $V=24 \; m/s$ airspeed from the \cite{dimitriadis2009bifurcation} experimental campaign, as this regime imposes the most rigorous numerical requirements.

\subsubsection{Grid Dependency Analysis}
\label{sssec:grid}
\begin{figure}
    \centering
    \includegraphics[width=1\linewidth]{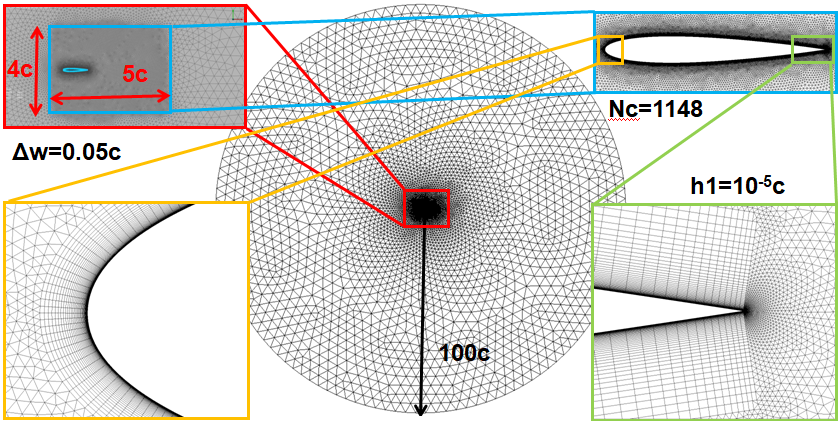}
    \caption{Overview of the O--type computational grid employed for the aeroelastic simulations. The domain extends to a far--field radius of $100c$ and totals approximately $105000$ cells. It consists of $N_c=1148$ cells along the airfoil perimeter, with a minimum discretisation length of $\Delta X_{min}=10^{-4}c$ at the leading and trailing edge and a growth rate of $1.1$ towards the maximum thickness region where a maximum length of $\Delta X_{max}=2*10^{-4}c$ is maintained. The boundary layer region comprises of $35$ cells with a first cell height of $h_1=10^{-5}c$ (ensuring $y^+ \leq 1$) and a growth ratio of $1.1$, whereas a refined near--wake region with a characteristic length of $\Delta w=5*10^{-2c}$ is used to capture the dynamic vortex shedding.}
    \label{fig:computational_grid}
\end{figure}

The initial focus of the sensitivity analysis is the refinement of the computational grid. An O--type mesh is utilised, extending to a far--field radius of one hundred chord lengths ($100c$) (see Figure \ref{fig:computational_grid}). A refined near--wake region is employed to properly resolve the emitted vortices, defined by a characteristic length of $\Delta w=5*10^{-2}c$. This region extends $5c$ in the chordwise direction ($0.5c$ upstream and $4.5c$ downstream) and $4c$ in the normal direction ($2c$ towards the pressure and $2c$ towards the suction side). For what is considered the medium grid, the airfoil surface is discretised in $N_c=1148$ cells along its perimeter, varying from a minimum spacing of $\Delta X_{min}=10^{-4}c$ at the leading and trailing edges to a maximum of $\Delta X_{max}=2*10^{-3}c$ close to the maximum thickness region, with an increase ratio of $1.1$. A structured mesh zone expands radially from the airfoil wall to ensure the correct prediction of the boundary layer development. The height of the first cell is $h_1=10^{-5}c$ (thereby ensuring a normalized wall distance of $y^+ \leq 1$) and the total number of cells in the radial direction is $35$ with a growth rate of $1.1$. For efficiency, an unstructured meshing technique is employed for the remainder of the domain, with a growth rate of $1.1$, resulting in a total number of approximately $105000$ cells.

A grid dependency analysis has been performed across four levels of refinement (Coarse, Medium, Fine and Extra Fine) to assess the variation in the predicted LCO characteristics (Amplitude and Period) listed in Table \ref{tab:grid}. All tested grids maintain a fixed minimum spacing of $\Delta X_{min}=10^{-4}c$ at the leading and trailing edges but differ in the maximum spacing ($\Delta X_{max}$), the number of cells defining the airfoil ($N_c$), and the characteristic length in the near--wake region ($\Delta w$). The Coarse grid ($\approx60000$ cells) uses $N_c=658$ cells to describe the airfoil shape, with a maximum spacing of $\Delta X_{max}=4*10^{-3}c$ and a characteristic length of $\Delta w=10^{-1}c$ in the near wake region. It demonstrates significant inaccuracy, underestimating the Pitch Amplitude by approximately $-9.4\%$ and overestimating the Period by approximately $+19.4\%$ relative to the Extra Fine mesh ($\approx440,000$ cells) that uses $N_c=2130$ cells in the airfoil wall, with a maximum spacing of $\Delta X_{max}=10^{-3}c$ and a characteristic length of $\Delta w=1.25*10^{-2}c$ in the near wake region. The Medium grid ($\approx105000$ cells) uses $N_c=1148$ cells to describe the airfoil shape, with a maximum spacing of $\Delta X_{max}=2*10^{-3}c$ and a characteristic length of $\Delta w=5*10^{-2}c$ in the near wake region and accomplishes substantially improved convergence, with variations of only approximately $+0.1\%$ in Amplitude and $+3.5\%$ in Period. The Fine mesh ($\approx215000$ cells) uses $N_c=2130$ cells in the airfoil wall, with a maximum spacing of $\Delta X_{max}=10^{-3}c$ and a characteristic length of $\Delta w=2.5*10^{-2}c$ in the near wake region and predicts similar deviations of approximately $-0.7\%$ and $+4.5\%$ in Amplitude and Period respectively. Consequently, the Medium mesh has been selected as the optimal compromise, providing accurate results at a manageable computational cost.

\begin{table}
\caption{Variation of Amplitude $[^{\circ}]$ and Period [sec] with grid refinement. Reference Amplitude and Period correspond to Extra Fine mesh comprising of $\approx 440000$ cells. The Medium mesh ($\approx 105000$ cells) provides the best compromise between accuracy and computational cost.}
\label{tab:grid}
    \centering
    \begin{tabular}{c|c|cc|cc}
    mesh type & $\# \: cells$ & \multicolumn{2}{c|}{Amplitude $[^{\circ}]$} & \multicolumn{2}{c}{Period [sec]}\\
    \hline
         Coarse & $\approx 60k$ & $21.31^{\circ}$ & $(-9.4\%)$ & $1.3175$ & $(+19.4\%)$\\
         Medium & $\approx 105k$ & $23.54^{\circ}$ & $(+0.1\%)$ & $1.1423$ & $(+3.5\%)$\\
         Fine & $\approx 215k$ & $23.36^{\circ}$ & $(-0.7\%)$ & $1.1535$ & $(+4.5\%)$\\
         Extra Fine & $\approx 440k$ & $23.52^{\circ}$ & $(-)$ & $1.1037$ & $(-)$\\
    \end{tabular}
\end{table}

\begin{figure}
    \centering
    \begin{subfigure}[t]{0.49\linewidth}
        \centering
        \includegraphics[width=\linewidth]{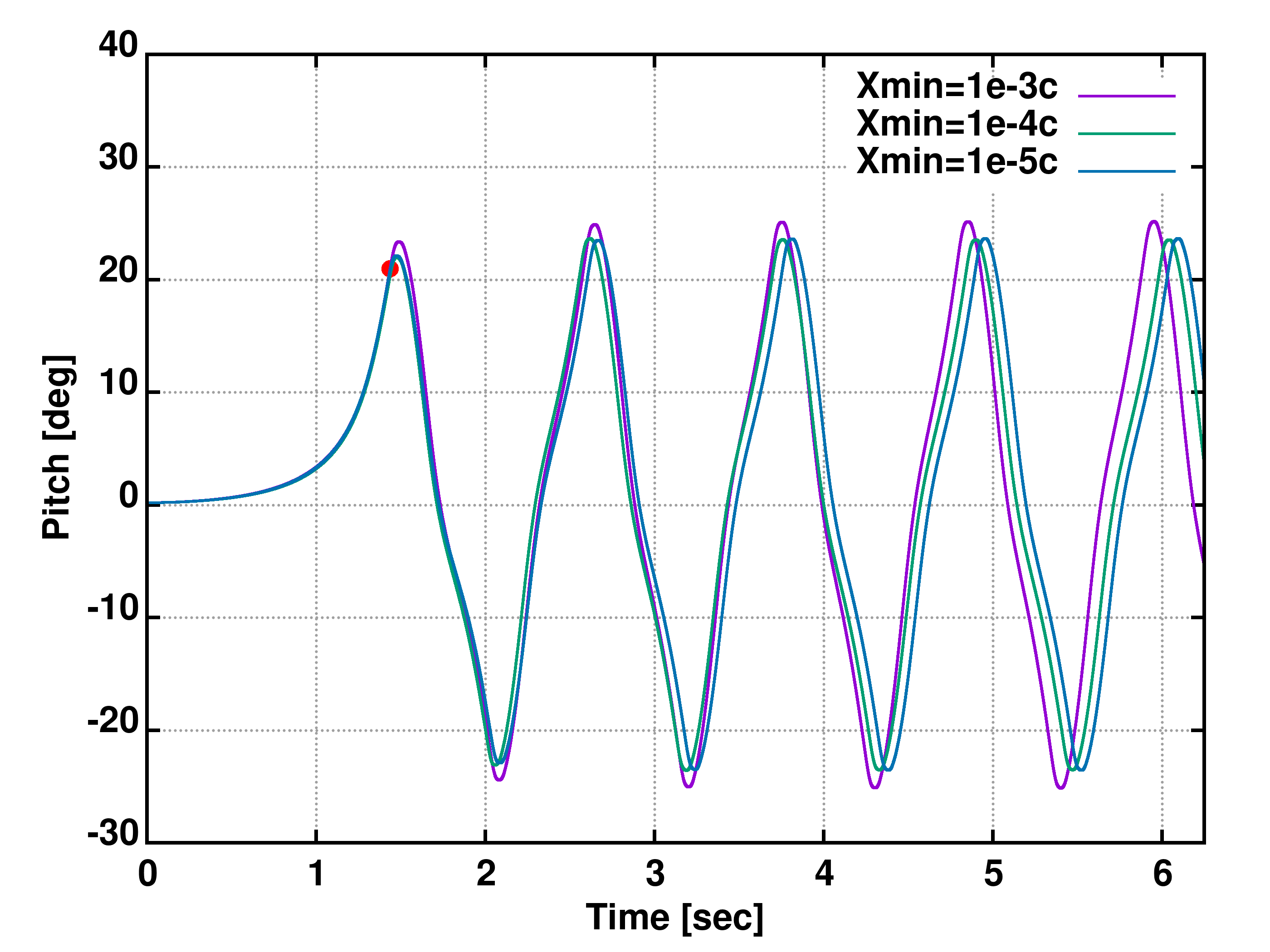}
        \caption{Pitch time history}
        \label{subfig:pitch_vs_time_Xmin}
    \end{subfigure}
    \hfill
    \begin{subfigure}[t]{0.49\linewidth}
        \centering
        \includegraphics[width=\linewidth]{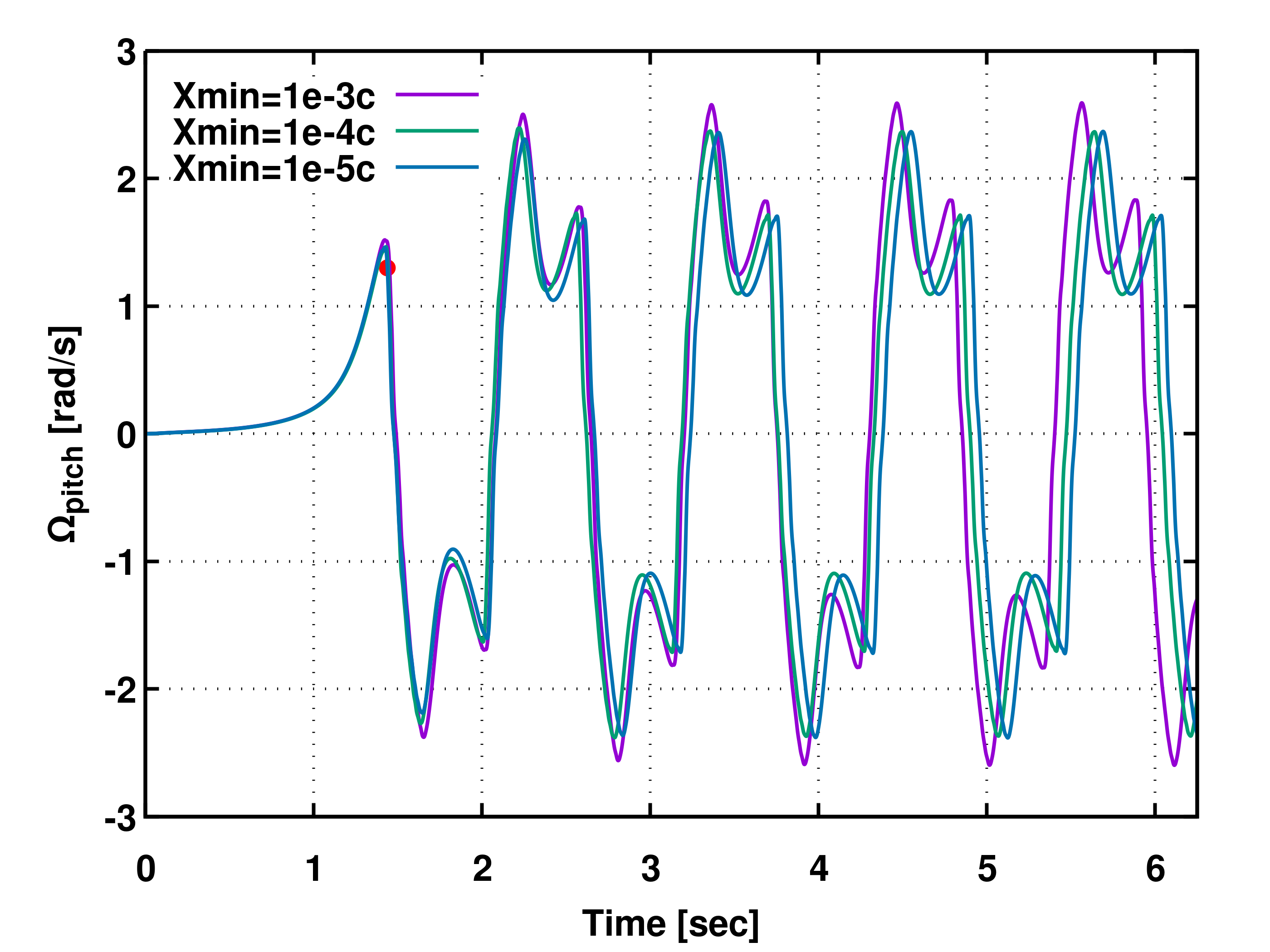}
        \caption{Pitch rate time history}
        \label{subfig:dpitch_vs_time_Xmin}
    \end{subfigure}
    \caption{Impact of minimum spacing ($\Delta X_{min}$) at the leading and trailing edges on pitch angle (a) and pitch rate (b) time history. All three simulations have a similar pitch angle and pitch rate (see Table \ref{tab:Xmin_pitch_dpitch}) at the specific time instant denoted by the red circle. Consequently, the deviations in the predicted pitch peak after the red dot can be attributed to different aerodynamic excitation at this time instant.}
    \label{fig:pitch_dpitch_vs_time_Xmin}
\end{figure}

\begin{figure}
    \centering
    \begin{subfigure}[t]{0.46\linewidth}
        \centering
        \includegraphics[width=0.9\linewidth]{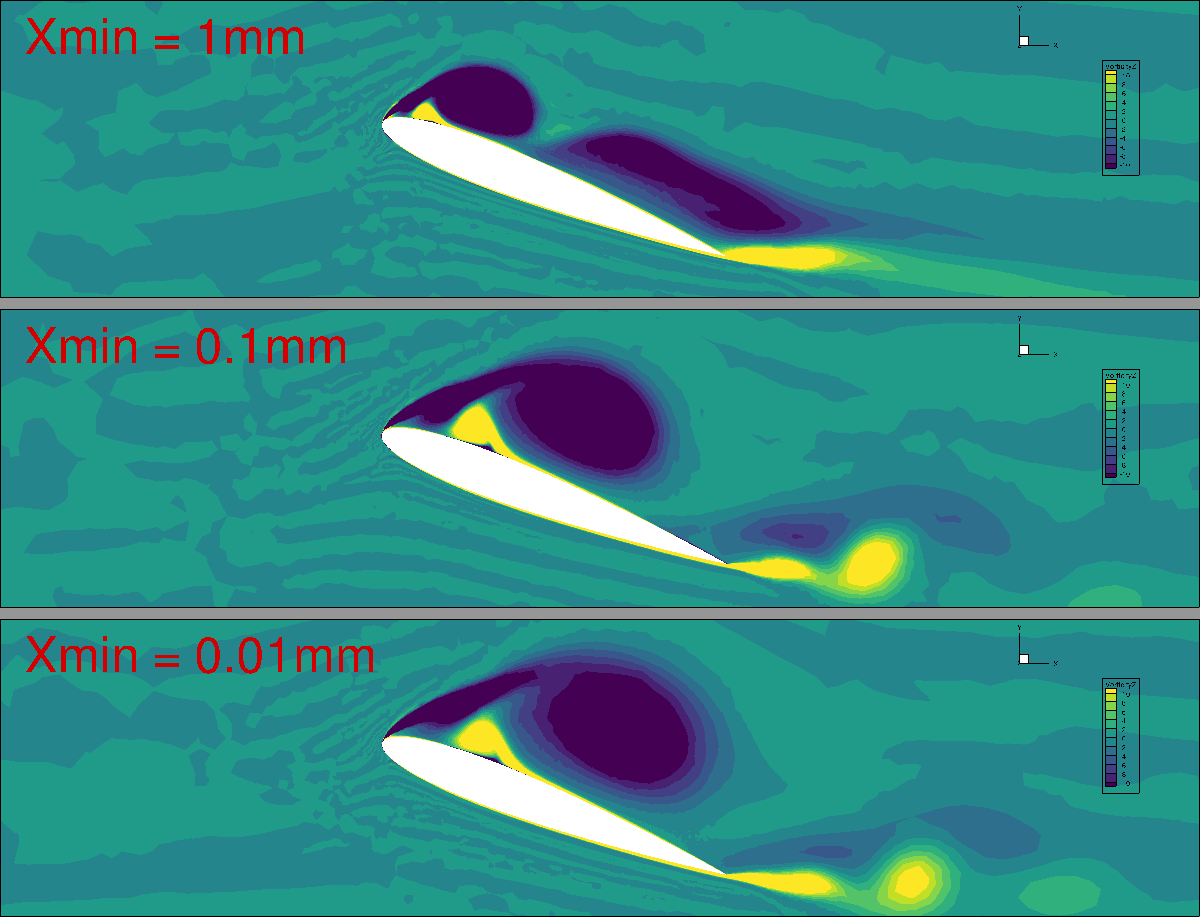}
        \caption{Vorticity controur}
        \label{subfig:vorticity_Xmin}
    \end{subfigure}
    \hfill
    \begin{subfigure}[t]{0.52\linewidth}
        \centering
        \includegraphics[width=0.9\linewidth]{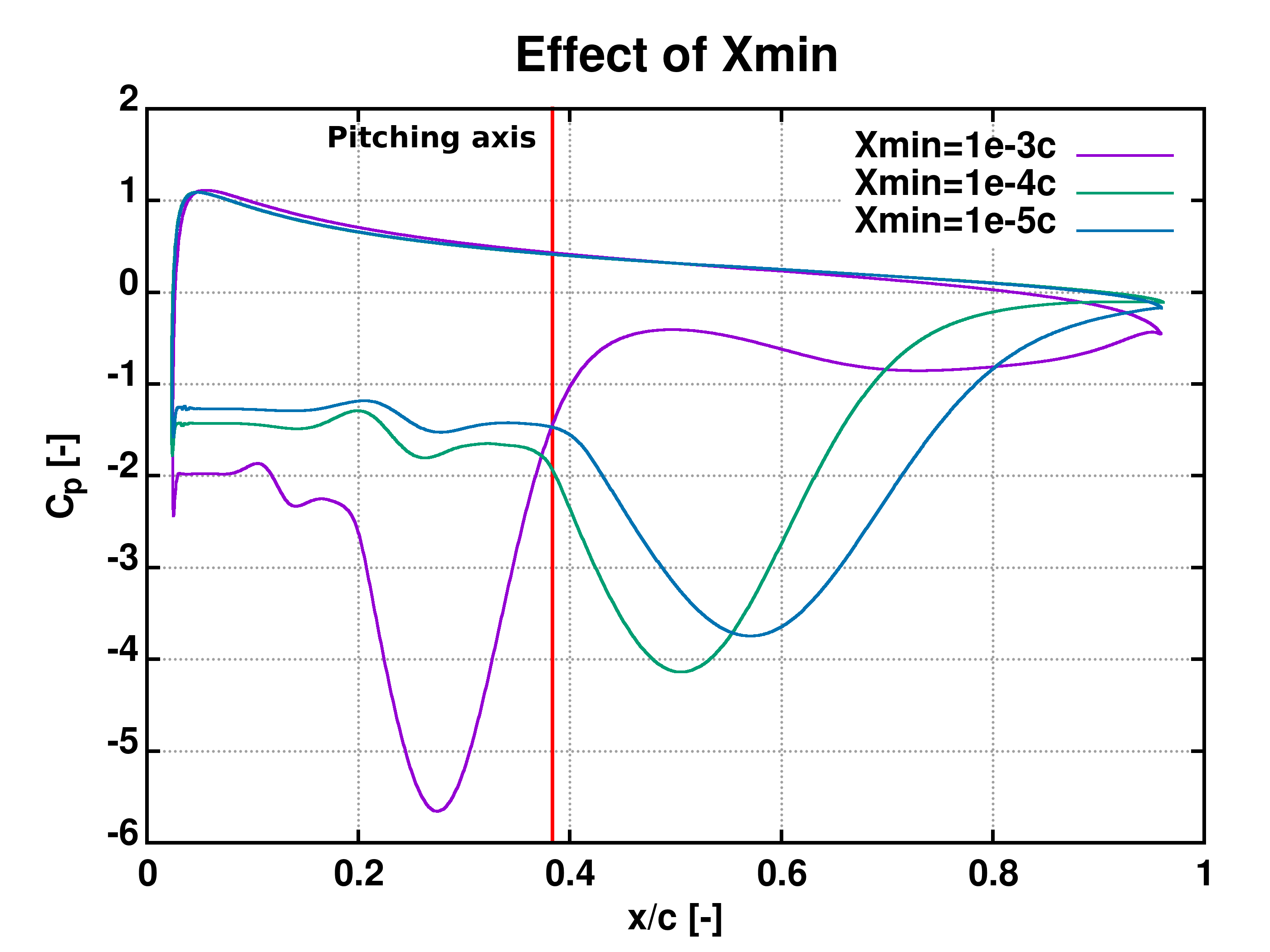}
        \caption{$C_p$ distribution}
        \label{subfig:cp_Xmin}
    \end{subfigure}
    \caption{Vorticity contour (a) and pressure coefficient distribution (b) for different leading edge and trailing edge discretisation lengths generated at the time instant (noted with a red dot in Figure \ref{fig:pitch_dpitch_vs_time_Xmin}). The influence of leading edge discretisation on leading edge vortex dynamics is highlighted, indicating that coarser discretisation hinders the detachment of the vortex, causing it to remain closer to the airfoil surface and resulting in an artificially increased excitation aerodynamic moment and thus larger predicted LCO amplitude.}
    \label{fig:Xmin}
\end{figure}

A specific analysis on the minimum spacing, $\Delta X_{min}$, has been also conducted, highlighting its critical influence on aeroelastic predictions. Table \ref{tab:Xmin} provides a clear overview of the effect of $\Delta X_{min}$ on the Amplitude and Period of the predicted LCO, clearly indicating that $\Delta X_{min}=10^{-4}c$ yields effectively in a grid--independent solution. On the other hand, when the coarse leading and trailing edge discretisation ($\Delta X_{min}=10^{-3}c$) is employed, Pitch Amplitude is overestimated by approximately $+6.7\%$ and the LCO Period is underestimated by approximately $-3.7\%$. Figure \ref{fig:pitch_dpitch_vs_time_Xmin} further illustrates this by showing the time history predicted when different values of $\Delta X_{min}$ are employed, where it is shown that the size of $\Delta X_{min}$ clearly affects the Amplitude and Period of the predicted pitching motion. This sensitivity stems from the influence of the grid resolution, particularly at the leading edge, on the precise development and subsequent detachment dynamics of the Leading Edge Vortex (LEV). Crucially, the flow field and pressure distribution over the airfoil presented in Figure \ref{fig:Xmin} are captured at the time instant denoted by the red circle in Figure \ref{fig:pitch_dpitch_vs_time_Xmin}. At this specific time, the differences observed in the aerodynamic field and thus the resulting aerodynamic moment are mainly attributed to the $\Delta X_{min}$ selection, as the airfoil exhibits similar pitch angle (see Figure \ref{subfig:pitch_vs_time_Xmin}) and angular velocity (see Figure \ref{subfig:dpitch_vs_time_Xmin}) across the tested discretisations up to this point (see Table \ref{tab:Xmin_pitch_dpitch}). When the leading edge discretisation is coarser, the LEV finds it challenging to detach, remaining closer to the airfoil (see Figure \ref{subfig:vorticity_Xmin}). This proximity leads to higher local velocities and thus increased suction aft from the pitching axis (see Figure \ref{subfig:cp_Xmin}). Ultimately, this leads to an increased excitation aerodynamic moment at this time instant that later results in higher pitch angles (see the different peak values predicted in \ref{fig:pitch_dpitch_vs_time_Xmin} right after the red dot). Conversely, finer discretisation facilitates smoother LEV shedding and convection, moving the vortex farther from the airfoil and thereby generating an increased restoring aerodynamic moment that drives the airfoil to lower pitch peak values after the red dot time instant. A similar qualitative outcome can be made for the trailing edge spacing. In fine TE discretisation, the TE shear layer has rolled up into a discrete detached vortex and a second one is already formed. For the crude dicretization the detachment of the TE vortex is delayed. These complex, non--trivial dependencies, visually supported by the flow field structures presented in Figure \ref{fig:Xmin}, demonstrate conclusively that grid dependency analyses on Stall Flutter instabilities must be extended beyond standard criteria, focusing specifically on how dynamic flow phenomena like vortex shedding are affected, rather than merely relying on static airfoil polar predictions.

\begin{table}
\caption{Variation of Amplitude $[^{\circ}]$ and Period [sec] with decreasing Leading Edge and Trailing Edge spacing $Xmin \: [m]$. Reference Amplitude and Period correspond to minimum value $Xmin = 10^-5c$. $Xmin = 10^-4c$ yields in a grid--independent solution.}
\label{tab:Xmin}
    \centering
    \begin{tabular}{c|cc|cc}
    $Xmin \: [-]$ & \multicolumn{2}{c|}{Amplitude $[^{\circ}]$} & \multicolumn{2}{c}{Period [sec]}\\
    \hline
         $10^-3c$ & $25.16^{\circ}$ & $(+6.7\%)$ & $1.0982$ & $(-3.7\%)$\\
         $10^-4c$ & $23.54^{\circ}$ & $(-0.2\%)$ & $1.1423$ & $(+0.1\%)$\\
         $10^-5c$ & $23.58^{\circ}$ & $(-)$ & $1.1408$ & $(-)$\\
    \end{tabular}
\end{table}

\begin{table}
\caption{Pitch angle and pitch rate values at the specific time instant denoted by the red circle in Figure \ref{fig:pitch_dpitch_vs_time_Xmin} for different values of minimum spacing ($\Delta X_{min}$) at the leading and trailing edges. All three simulations have a similar pitch angle and pitch rate at this instant.}
\label{tab:Xmin_pitch_dpitch}
    \centering
    \begin{tabular}{c|c|cc}
    $Xmin \: [-]$ & Pitch angle $[^{\circ}]$ & Pitch rate $[rad/s]$\\
    \hline
         $10^-3c$ & $20.98$ & $1.50$ \\
         $10^-4c$ & $20.38$ & $1.40$\\
         $10^-5c$ & $20.82$ & $1.31$\\
    \end{tabular}
\end{table}

\subsubsection{Time--step Dependency Analysis}
\label{sssec:time}

\begin{figure}
    \centering
    \begin{subfigure}[t]{0.49\linewidth}
        \centering
        \includegraphics[width=\linewidth]{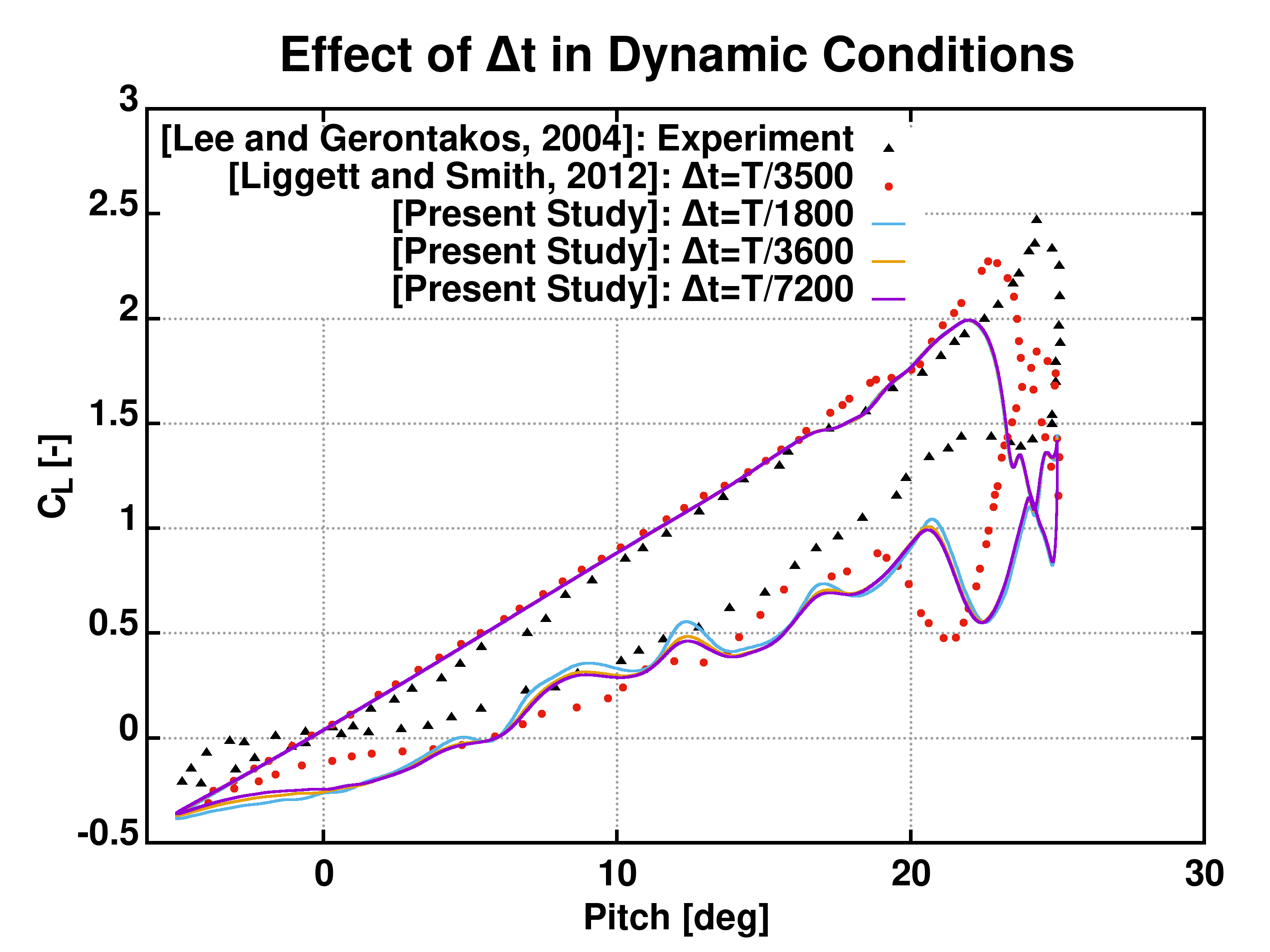}
    \caption{Dynamically pitching NACA0012 (imposed motion) at $a_0 = 10^{\circ}$, $a_1 = 15^{\circ}$ and $k = 0.1$, at $Re=135000$ and $Ma=0.103$. Experimental data from \cite{lee2004investigation} and numerical predictions from \cite{liggett2012temporal}.}
    \label{subfig:cl_dt_dynamic}
    \end{subfigure}
    \hfill
    \begin{subfigure}[t]{0.49\linewidth}
        \centering
        \includegraphics[width=\linewidth]{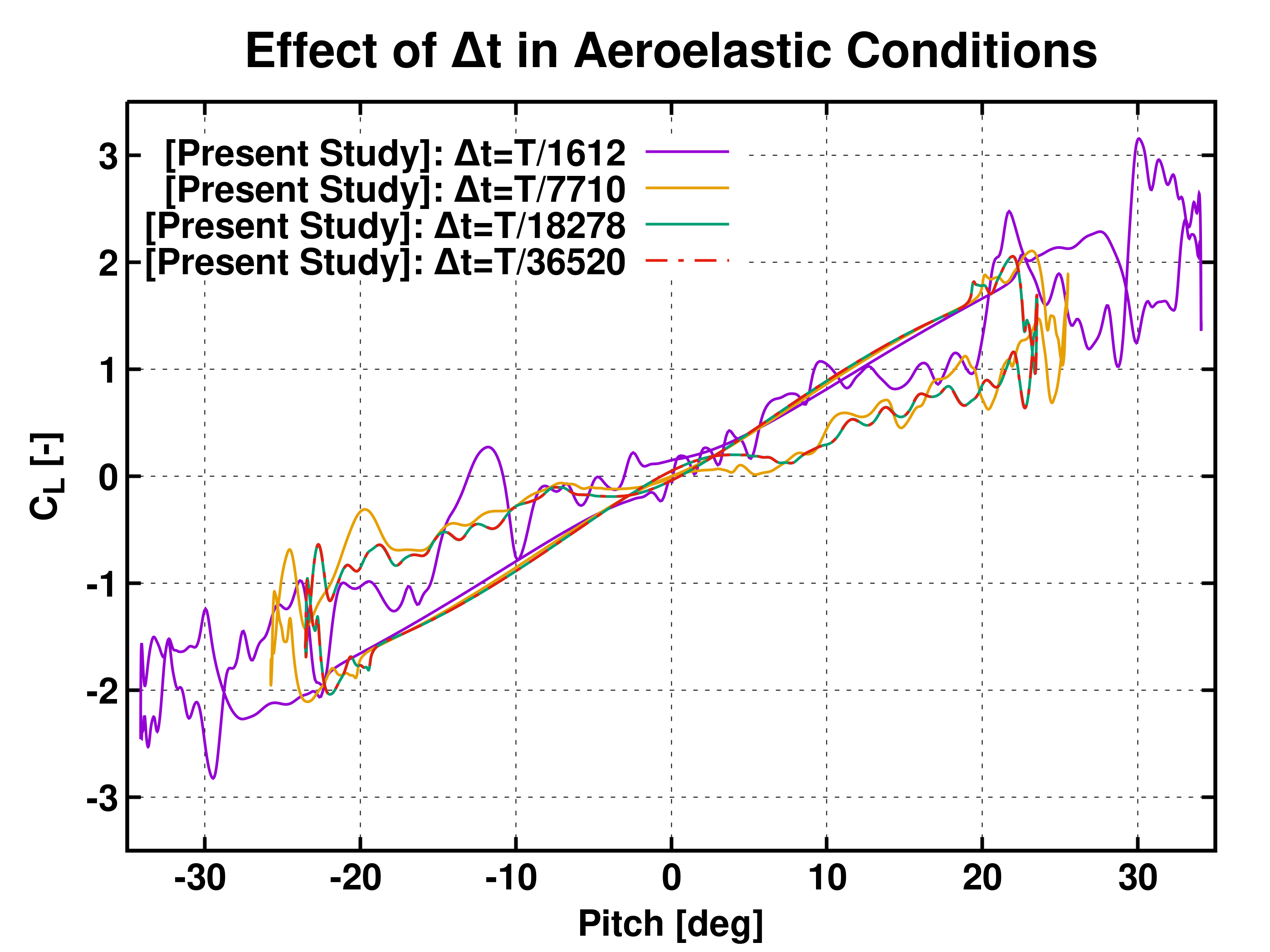}
        \caption{Predictions by MaPFlow for an aeroelastically pitching NACA0012 in symmetric LAO at $Re=540000$ and $Ma=0.079$, firstly observed in \cite{dimitriadis2009bifurcation} experimental campaign.}
        \label{subfig:cl_dt_aeroelastic}
    \end{subfigure}
    \caption{Lift coefficient variation with pitch angle in dynamic (a) and aeroelastic conditions (b) for a pitching NACA0012. The time--step value required to correctly resolve the aeroelastic response of a pitching airfoil (b) is significantly smaller compared to the one needed to accurately predict the aerodynamic loading of a rigid airfoil undergoing imposed pitching motion (a).}
    \label{fig:cl_dt_NTUA}
\end{figure}

Table \ref{tab:dt} shows how the selected non--dimensional time--step $\Delta \tau = \frac{\Delta t \: V}{c}$ affects the aeroelastic response of the system. Except from the value of $\Delta \tau=0.02$ that results in ever--increasing amplitude, a monotonic convergence is observed, with $\Delta \tau=0.005$ yielding a time--step independent solution. This non--dimensional time-step value ensures that at least $14400$ steps per oscillation period will be utilised to simulate the aeroelastic response of the system. Based on this outcome, the time--step value $\Delta t$ turns out to be yet another numerical parameter that shows increased sensitivity when investigated within an aeroelastic context. This is clearly shown in Figure \label{fig:cl_dt_NTUA}, where the impact of the time--step value in the predicted lift coefficient variation is depicted for different oscillating conditions. In Figure \ref{subfig:cl_dt_dynamic} it is shown that a $\Delta t$ corresponding to only $3600$ steps per period is found to be sufficient for a time--step independent prediction of a dynamically pitching airfoil, whereas more than $14400$ steps per period are necessary for the aeroelastic case shown in Figure \ref{subfig:cl_dt_aeroelastic}. This lenient requirement resulting from dynamic cases of rigid airfoils undergoing imposed pitching motion is considered as standard practice for time--step independence analyses found in literature. Nevertheless, it is clearly shown in this study that the time--step value required to correctly resolve the aeroelastic response of an airfoil undergoing stall flutter oscillations is significantly smaller.

\begin{table}
\caption{Variation of Amplitude $[^{\circ}]$ and Period [sec] with decreasing non--dimensional time--step value $\Delta \tau = \frac{\Delta t \: V}{c} [-]$. Reference Amplitude and Period correspond to finest value $\Delta \tau = 0.0025$. $\Delta \tau = 0.005$ yields in a time--step independent solution.}
\label{tab:dt}
    \centering
    \begin{tabular}{c|cc|cc}
    $\Delta \tau \: [-]$ & \multicolumn{2}{c|}{Amplitude $[^{\circ}]$} & \multicolumn{2}{c}{Period [sec]}\\
    \hline
         $0.04$ & $34.14$ & $(+45\%)$ & $0.7853$ & $(-31.2\%)$\\
         $0.02$ & $\infty$ & & $\infty$ &\\
         $0.01$ & $25.80$ & $(+9.6\%)$ & $0.971$ & $(-14.8\%)$\\
         $0.005$ & $23.5$ & $(+0\%)$ & $1.1423$ & $(+0.1\%)$\\
         $0.0025$ & $23.54$ & $(-)$ & $1.1412$ & $(-)$\\
    \end{tabular}
\end{table}

\section{Results}
\label{sec:results}
In this section, the numerical results are presented by applying the validated methodology and numerical setup presented in Section \ref{sec:methodology}. The investigation focuses on the two distinct Stall Flutter regimes: SAO in the transitional Reynolds number range and LAO in the moderate Reynolds number regime.

\subsection{Small Amplitude Oscillations}
\label{ssec:low_amplitude}
\begin{figure}
    \centering
    \includegraphics[width=0.85\linewidth]{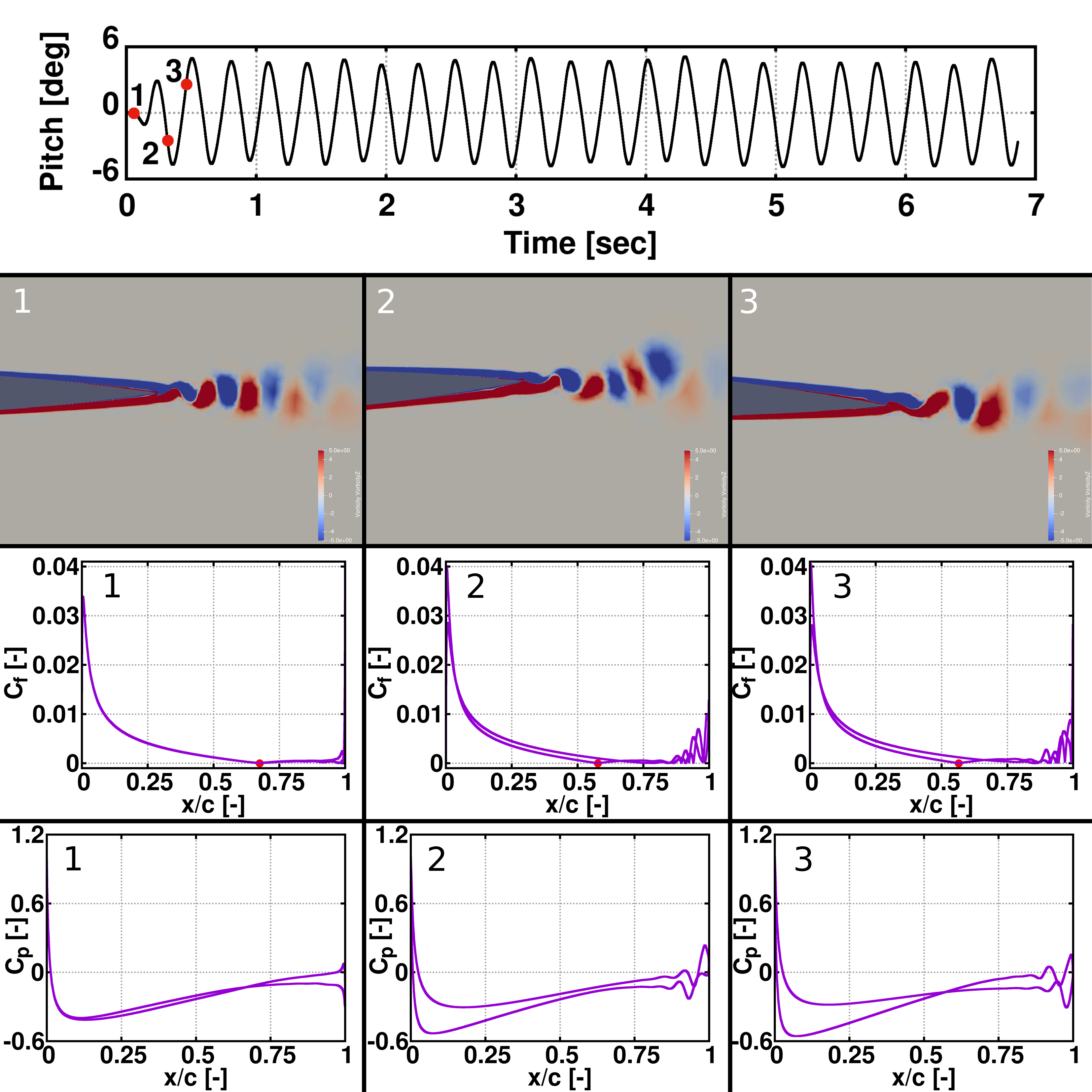}
    \caption{Pitch response time history, vorticity contours, friction and pressure coefficient distribution for a SAO case at a Reynolds number of $Re=1.17*10^5$, firstly observed in \cite{poirel2008self} experimental campaign. Three critical time instants have been makred in the time history plot, namely: "$1$" where the emergence of trailing edge vortices leads to an instability onset that drives the system to oscillations, "$2$ and "$3$" where the subsequent formation of larger detached vortices stabilizes the system to LCO. The corresponding friction and pressure coefficient distributions imprint the flow dynamics associated with the vortices formation. Red dots in friction coefficient plots indicate the laminar separation positions where LSBs form.}
    \label{fig:flowfield_snapshots_poirel}
\end{figure}

\begin{figure}
    \centering
    \includegraphics[width=0.85\linewidth]{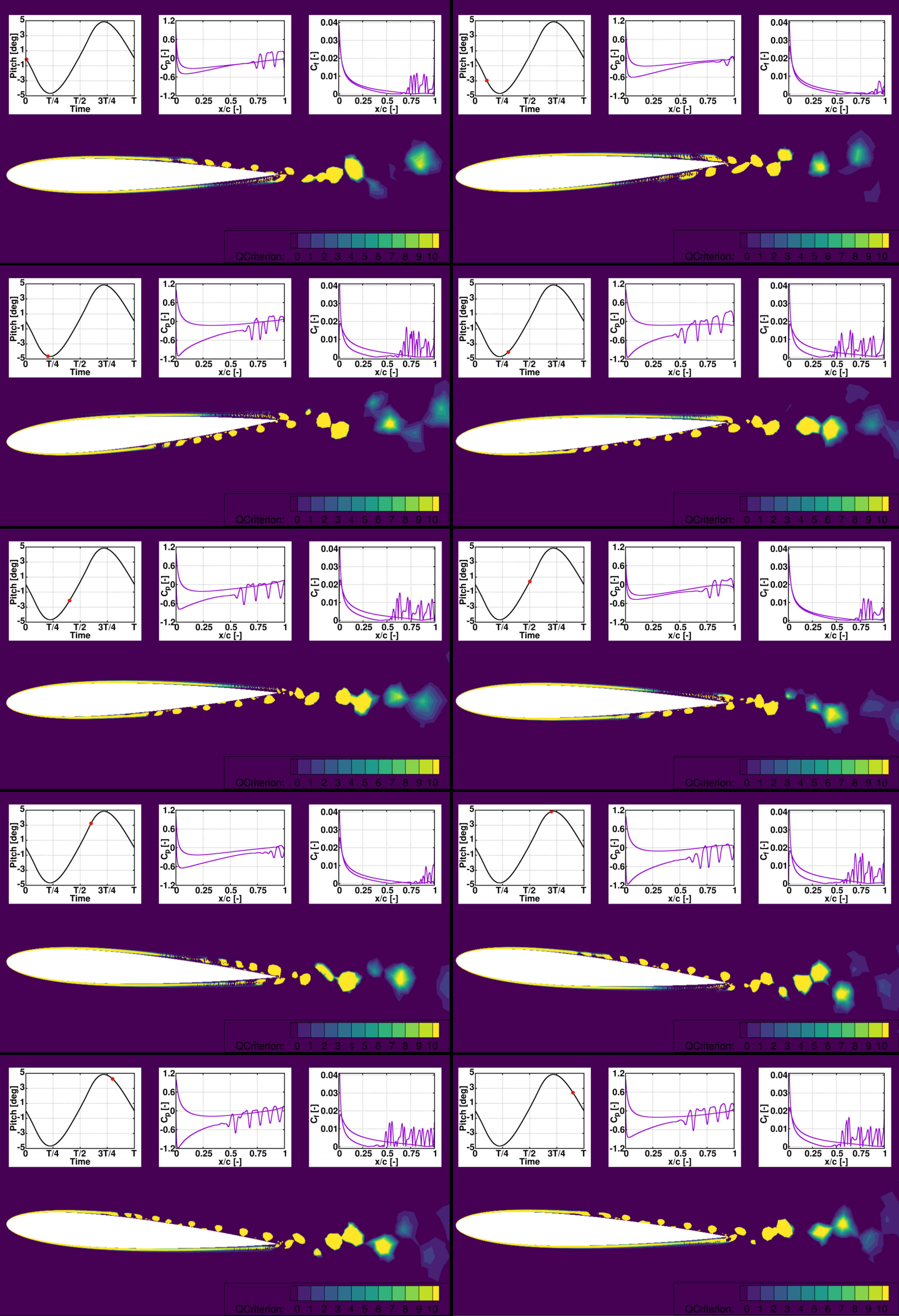}
    \caption{Pitch response time history, Q--criterion contours, friction and pressure coefficient distribution for a SAO case at a Reynolds number of $Re=1.17*10^5$, firstly observed in \cite{poirel2008self} experimental campaign. The presence of a LSB is computationally verified in $C_f$ plots, at the positions where the $C_f$ values drop to zero. Nevertheless, Q--criterion contours indicate a von Kármán vortex shedding as the LCO sustaining factor, forming upstream--propagating instabilities.}
    \label{fig:Q-criterion_snapshots_poirel}
\end{figure}

In this sub--section, the aeroelastic response of a system that results in SAO is examined. This phenomenon has been observed in a low and constrained transitional Reynolds number range (from $45000$ to $130000$). In \cite{poirel2008self} it is stated that the flow separation close to the trailing edge plays a determinant role in the onset of the fluttering motion, while the subsequent presence of a LSB stabilizes this motion into a LCO. Nevertheless, CFD predictions in \cite{poirel2011computational} indicate a Kelvin--Helmholtz instability occurring in the shear layer close to the trailing edge as the onset mechanism to the fluttering motion and the resulting von Kármán vortex shedding as the LCO stabilisation factor. MaPFlow predictions, illustrated in Figure \ref{fig:flowfield_snapshots_poirel}, confirm this mechanism, showing that the emergence of trailing edge shedding at time--stand "$1$" drives the system towards an instability onset, which is subsequently stabilized by the formation of larger detached vortices near the trailing edge (time--stands "$2$" and "$3$") that bring about the large unsteady fluctuations on the aft part of the airfoil shown in $C_f$ and $C_p$ diagrams. The presence of a LSB is also verified by the present study simulations in the positions where the wall shear stress vanishes (indicated by red dots). However, as shown in the Q--criterion contours displayed in Figure \ref{fig:Q-criterion_snapshots_poirel}, a von Kármán vortex shedding is the sustaining factor of the fluttering motion that induces upstream--propagating instabilities in both sides of the airfoil. The employment of transition modelling is considered essential to correctly capture the initial trailing edge separation, consequently leading to the SAO regime. Neglecting transition in the numerical simulations results in the complete suppression of the LCO.

\begin{figure}
    \centering
    \begin{subfigure}[t]{0.49\linewidth}
        \centering
        \includegraphics[width=0.9\linewidth]{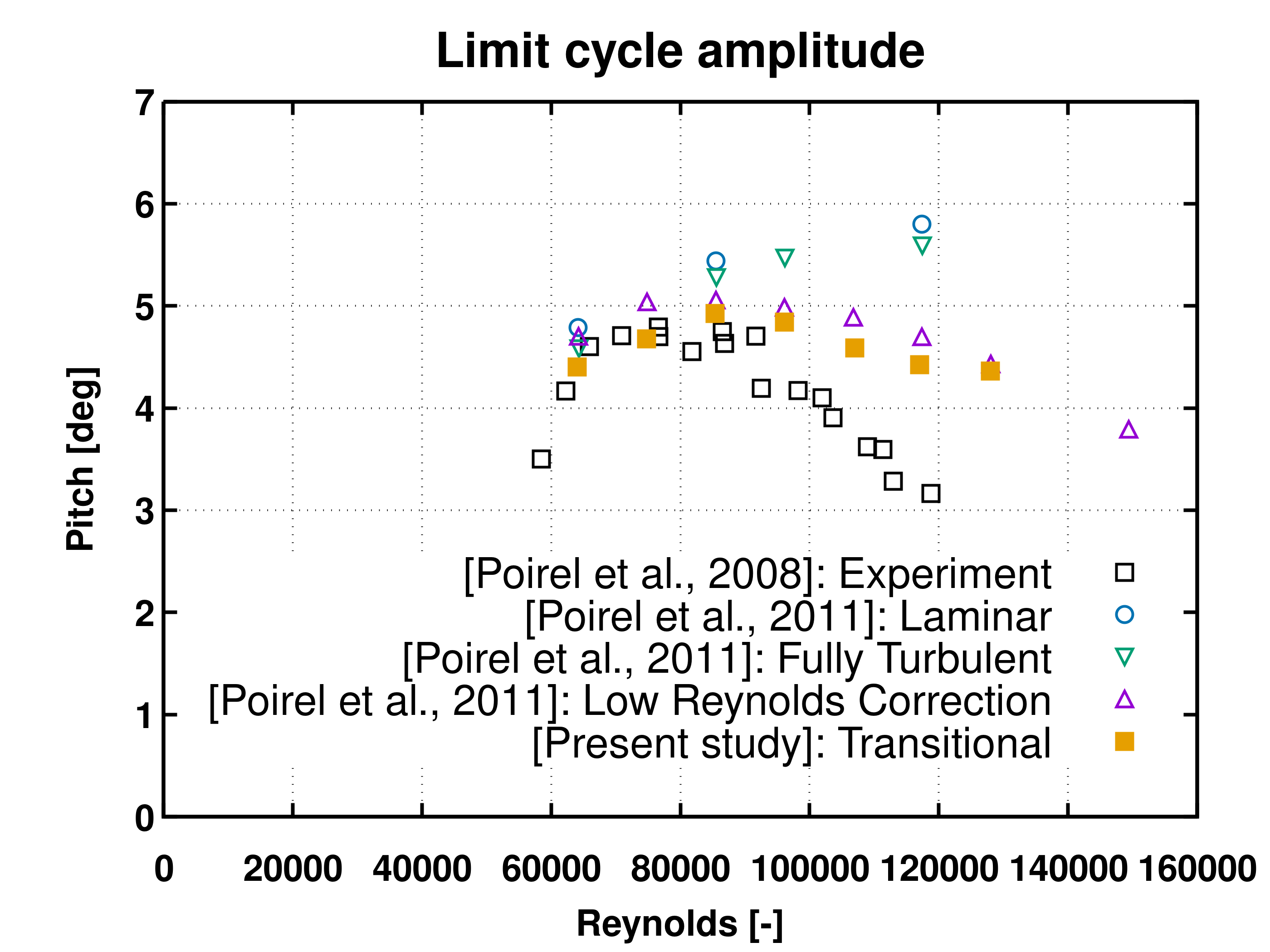}
        \caption{Pitch Amplitude}
        \label{subfig:amp_vs_re}
    \end{subfigure}
    \hfill
    \begin{subfigure}[t]{0.49\linewidth}
        \centering
        \includegraphics[width=0.9\linewidth]{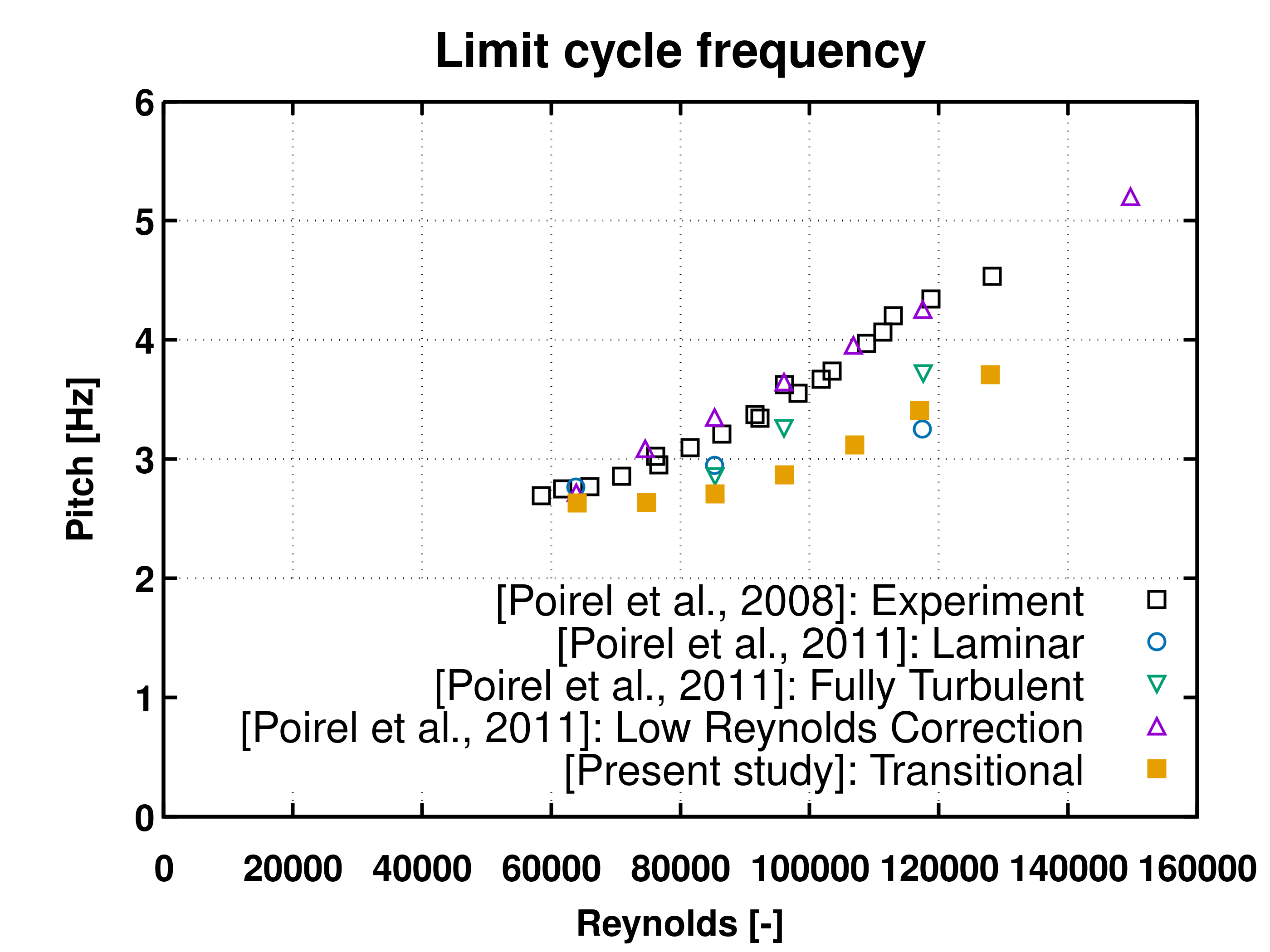}
        \caption{Pitch Frequency}
        \label{subfig:freq_vs_re}
    \end{subfigure}
    \caption{Pitch amplitude (left) and frequency (right) variation with respect to Reynolds number. Experimental measurements from \cite{poirel2008self} and predictions with different turbulence modelling parameters by other CFD tools from \cite{poirel2011computational}. The Low Reynolds Correction and the Transitional models can fairly capture the increase in the amplitude at low Reynolds numbers ($Re \leq 85400$) and the following decrease as the Reynolds number further increases, whereas other predictions show a significant deviation from the measured trend. The predicted amplitudes are slightly larger than the measured ones in the latter Reynolds range for both computational options, which accounts for the corresponding lower frequencies predicted by the Transitional model for the same Reynolds range. Nevertheless, the trend of the increasing frequency along with the Reynolds number is generally predicted with accuracy.}
    \label{fig:reynolds_pass}
\end{figure}

Figure \ref{fig:reynolds_pass} presents a comprehensive comparison of the resulting LCO characteristics, specifically amplitude and frequency, for various Reynolds numbers in the transitional range. Experimental measurements are juxtaposed against numerical predictions obtained using various turbulence modelling options. It is evident that Laminar and the standard Fully Turbulent simulations deviate significantly from the experimentally measured trend in amplitude (see Figure \ref{subfig:amp_vs_re}). However, the non--monotonic trend of increasing amplitude at lower Reynolds numbers ($Re \leq 85400$) followed by a decrease as the Reynolds number further increases, is successfully captured by the Low Reynolds Correction and the Transitional simulations. Nevertheless, a persistent over--prediction of amplitudes in the higher Reynolds range is shown, where computational models predictions agree well with each other. This amplitude overestimation directly correlates with the prediction of lower oscillation frequencies by the present study simulations in the same range (see Figure \ref{subfig:freq_vs_re}). However, the overall trend of increasing oscillation frequency with increasing Reynolds number is fairly predicted.

\subsection{Large Amplitude Oscillations}
\label{ssec:large_amplitude}
In this sub--section, the aeroelastic response resulting in LAO is examined. This phenomenon has been observed by \cite{li2007experimental} in a moderate Reynolds number regime ranging from $2.4*10^5$ up to $6*10^5$ ($12$ to $30 \; m/s$). Unlike SAO which are associated with laminar separation flutter in the transitional regime, LAO are fundamentally linked to deep dynamic stall caused by the generation and subsequent shedding of a strong LEV. Consequently, accurate prediction of these LAO oscillations across the full experimental range necessitates fully turbulent simulations using the standard k--$\omega$ SST turbulence model.

\begin{figure}
    \centering
    \includegraphics[width=0.65\linewidth]{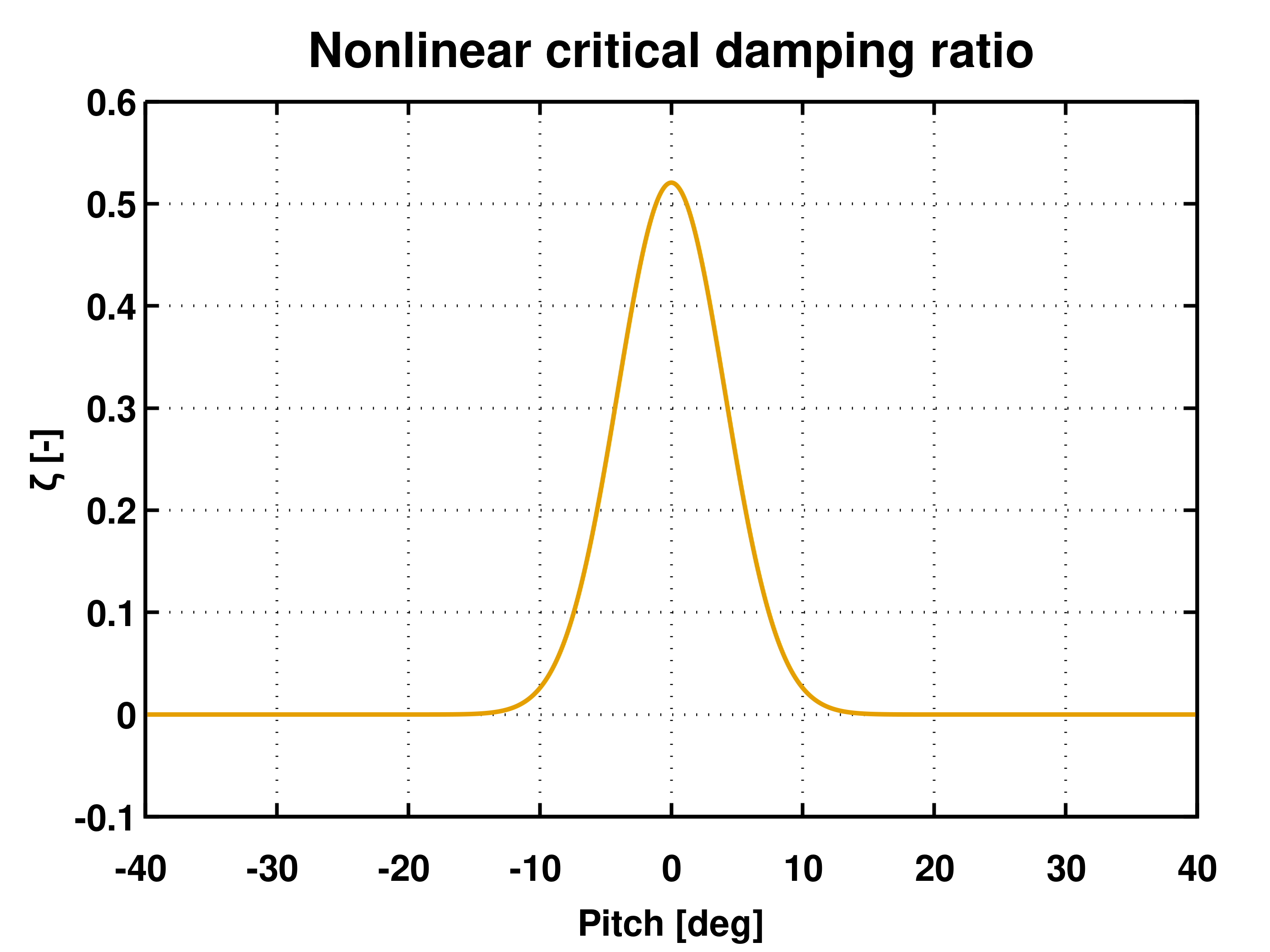}
    \caption{Non--linear critical damping ratio $\zeta$ with respect to pitch angle $\theta$ (in degrees) employed in LAO simulations of \cite{dimitriadis2009bifurcation} cases.}
    \label{fig:non_linear_damping}
\end{figure}

The structural damping in the experimental setup (primarily due to friction in the pitch springs) was found to exhibit highly non--linear characteristics. Experimental observations, specifically mentioned in Section \ref{subsec:LAO_exp_setup}, indicated that the critical pitch damping coefficient $\zeta$ demonstrated pronounced non--linearity primarily near small pitch angles (around the zero position) and transitioned to a linear increase at higher pitch angles. This behavior necessitates the implementation of a non--linear damping model, typically structured as a function of the pitch angle $\zeta(\theta)$, as describe by Equation \eqref{eq:LAO_zeta} and illustrated in Figure \ref{fig:non_linear_damping}. 

\begin{equation}
\label{eq:LAO_zeta}
    \begin{split}
    \zeta(\theta) &= 0.52e^{-0.03\theta^2} \\
    C(\theta) &= 2\zeta(\theta)\sqrt{IK}
    \end{split}
\end{equation}

Although the damping was noted to increase linearly at larger pitch angles, the component representing this linear increase has been entirely omitted in the final model. This simplification is justified by the fact that the corresponding angular velocity $\dot{\theta}$ is typically small at large angles, meaning that the contribution of this linear term to the overall off--wind response is insignificant. Omitting this term consequently reduces the difficulty of tuning the model by removing an additional unknown variable. Figure \ref{fig:free_decay} presents the successful validation of the chosen non--linear damping function of Equation \eqref{eq:LAO_zeta} against experimental free--decay tests Do these additional comments suffice?]{performed at an airspeed of $0$, $2.5$ and $3.1\;m/s$. The complete aeroelastic system has been analysed in these simulations. Aeroodynamics has been properly accounted for, even though the airflow velocity is very low, meaning that the only significant flow component comes from the pitching motion of the wing. Deflection from the initial resting position has been accomplished by applying an impulse, which can be modelled either as an initial non--zero pitch angle (see Figures \ref{subfig:free_decay_0} and \ref{subfig:free_decay_2.5}) or a sudden moment around the pitching axis (see Figure \ref{subfig:free_decay_3.1}).} The comparison confirms that this model accurately captures the dynamic response of the system in the absence of significant aerodynamic excitation, illustrating the characteristic rapid decay of oscillations observed experimentally.
\begin{figure}
    \centering
    \begin{subfigure}[t]{0.65\linewidth}
        \centering
        \includegraphics[width=\linewidth]{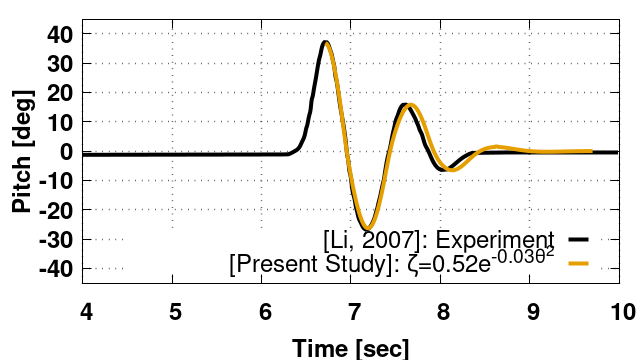}
        \caption{$V=0 \; m/s$}
        \label{subfig:free_decay_0}
    \end{subfigure}
    \vfill
    \begin{subfigure}[t]{0.65\linewidth}
        \centering
        \includegraphics[width=\linewidth]{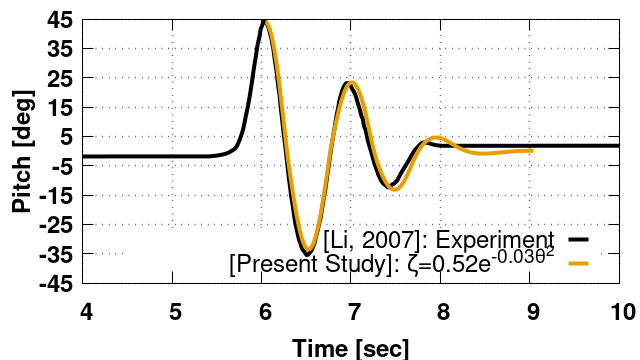}
        \caption{$V=2.5 \; m/s$}
        \label{subfig:free_decay_2.5}
    \end{subfigure}
    \label{fig:free_decay}
\end{figure}

\begin{figure}
    \ContinuedFloat
    \centering
    \begin{subfigure}[t]{0.65\linewidth}
        \centering
        \includegraphics[width=\linewidth]{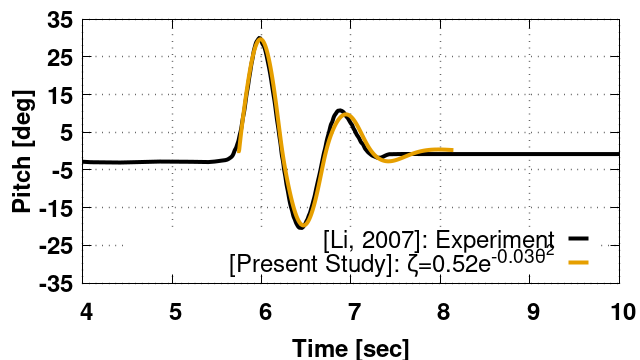}
        \caption{$V=3.1 \; m/s$}
        \label{subfig:free_decay_3.1}
    \end{subfigure}
    \caption{Numerical prediction of the free--decay tests performed at $0$ (a), $2.5$ (b) and $3.1$ (c) $m/s$ airflow speed. The comparison against experimental measurements confirms that the non--linear damping model described in Equation \eqref{eq:LAO_zeta} accurately captures the dynamic response of the system in the absence of significant aerodynamic excitation.}
    \label{fig:free_decay}
\end{figure}

\begin{figure}
    \centering
    \begin{subfigure}[t]{0.49\linewidth}
        \centering
        \includegraphics[width=\linewidth]{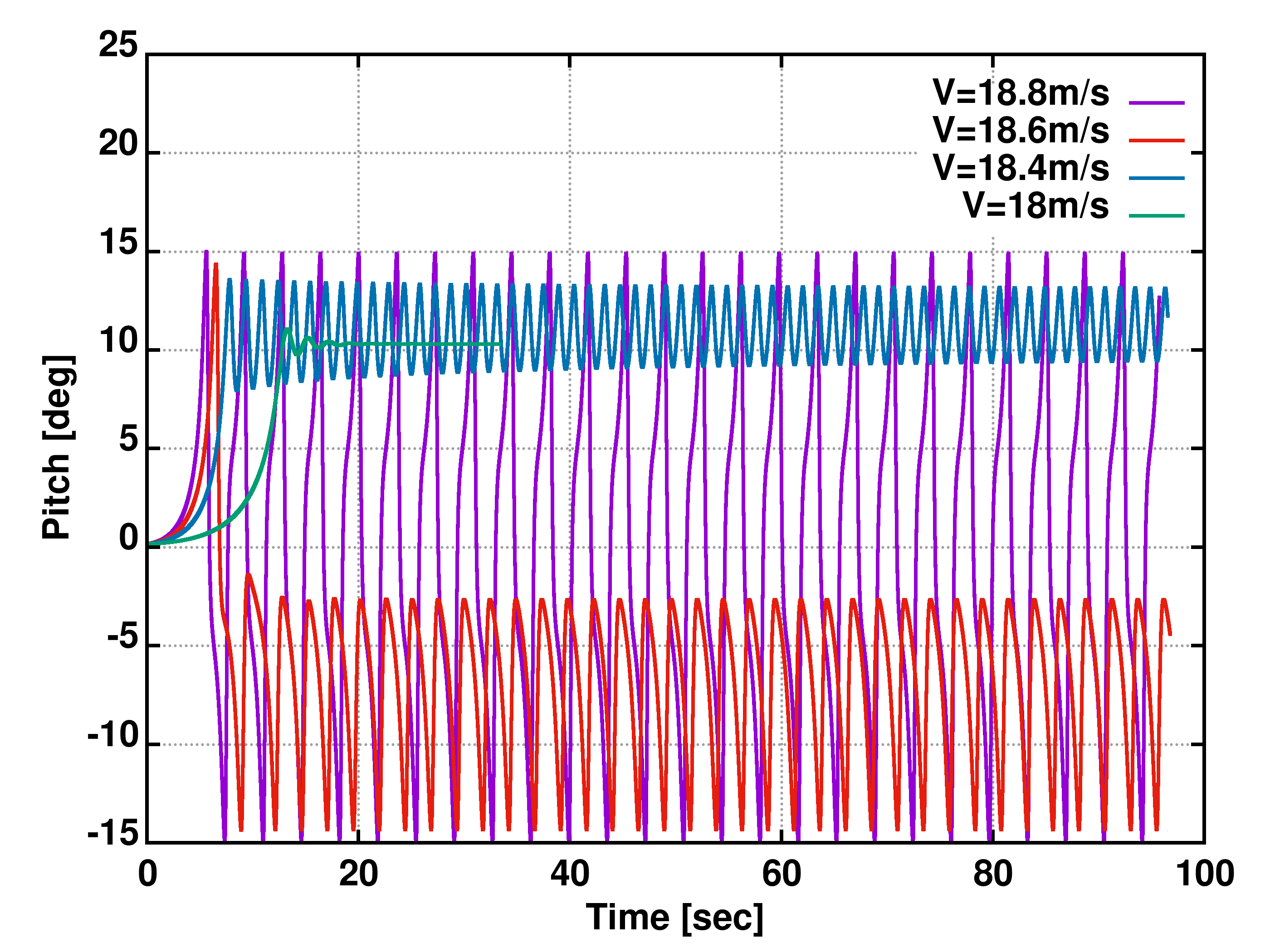}
    \label{subfig:pitch_vs_time_V19ms}
    \end{subfigure}
    \hfill
    \begin{subfigure}[t]{0.49\linewidth}
        \centering
        \includegraphics[width=\linewidth]{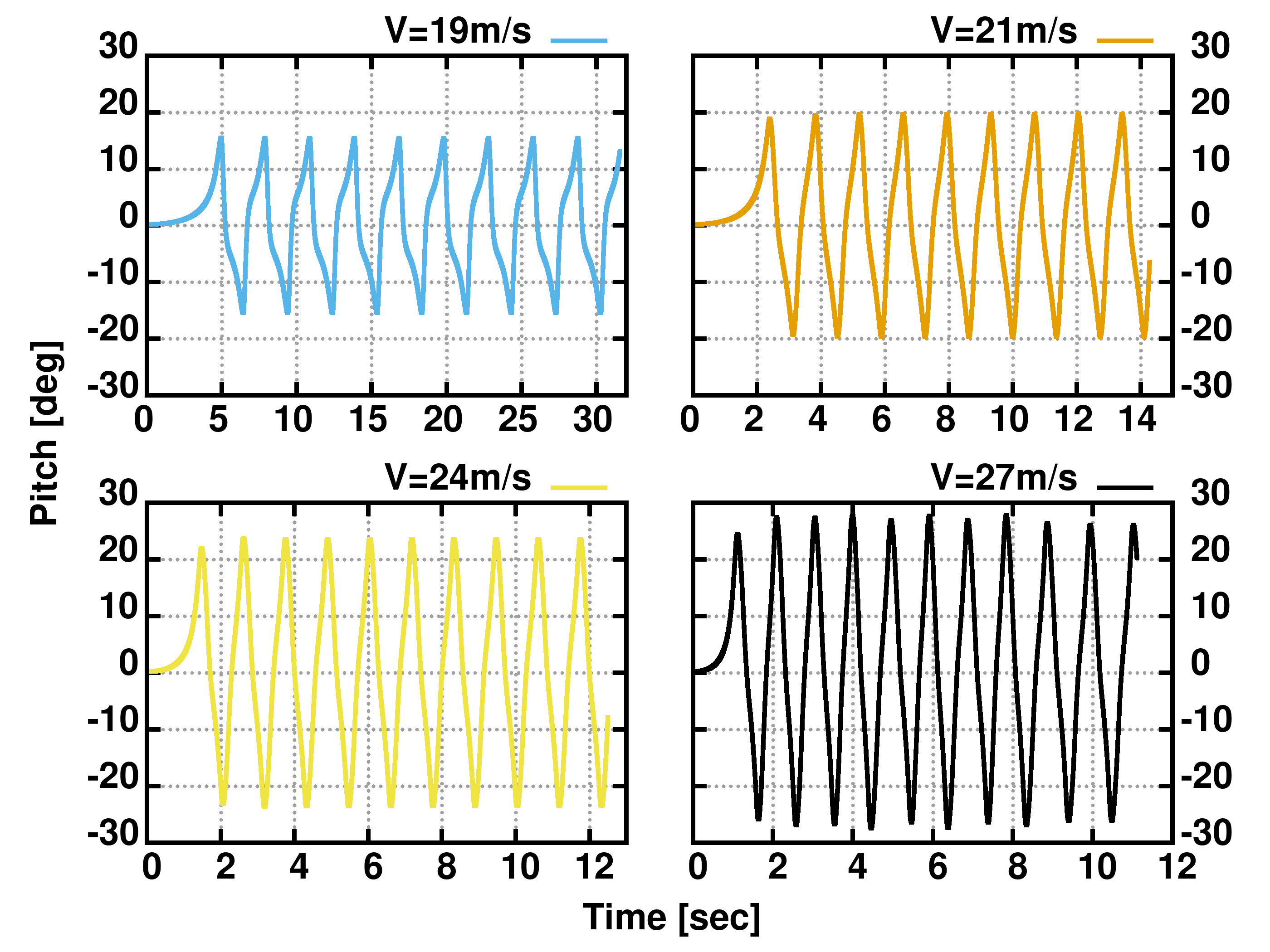}
        \label{subfig:pitch_vs_time_V19ms}
    \end{subfigure}
    \caption{Pitch angle time history for various airflow velocities demonstrating the transition from static stability to LCO. Cases shown include stable decay to the static equilibrium position ($V=18 \; m/s$), marginally stable convergence ($V=18.4 \; m/s$) and the onset of self--started and sustained oscillations at the numerical critical velocity ($V_{c,num}=18.6 \; m/s$) where asymmetric oscillations around the respective static equilibrium angle are predicted along with a bifurcation from the symmetric to the asymmetric aeroelastic mode. For higher airflow velocities, only symmetric oscillations of significantly larger amplitude around the zero mean pitch angle are predicted.}
    \label{fig:pitch_vs_time}
\end{figure}

\begin{figure}
    \centering
    \includegraphics[width=0.7\linewidth]{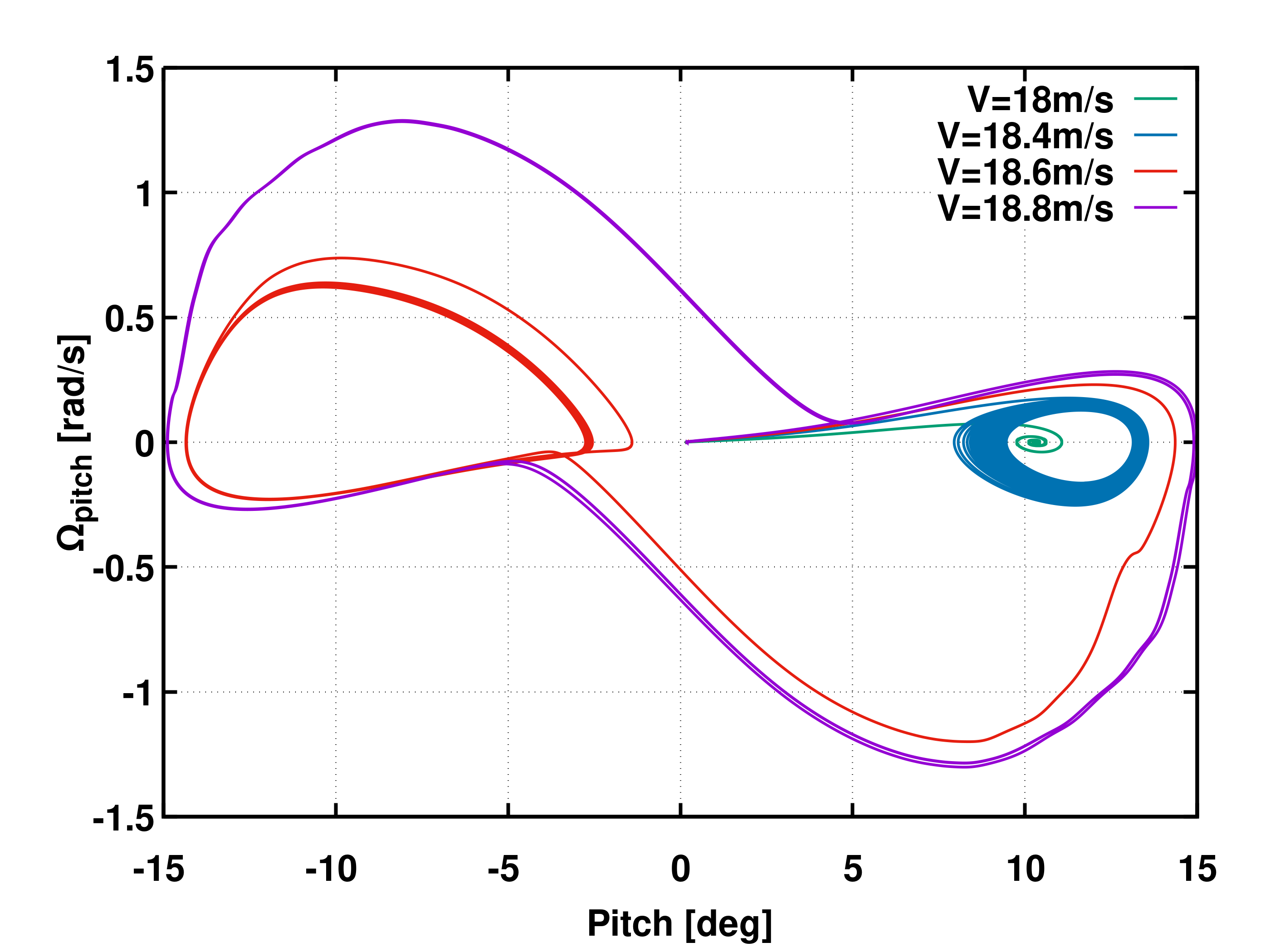}
    \caption{Phase plane trajectories ($\theta$ versus $\dot{\theta}$) for various airflow velocities. Trajectories depict stable convergence to the static equilibrium point ($V=18 \; m/s$), marginally stable convergence ($V=18.4 \; m/s$), small amplitude asymmetric oscillations near the critical velocity ($V=18.6 \; m/s$), and large amplitude symmetric oscillations around zero mean pitch angle ($V=18.8 \; m/s$).}
    \label{fig:pitch_vs_dpitch}
\end{figure}

Figure \ref{fig:pitch_vs_time} and Figure \ref{fig:pitch_vs_dpitch} illustrate the transition from static stability to LCO for increasing values airflow velocity. When the aeroelastic system is exposed to low airflow velocity, such as $V=18 \; m/s$, the airfoil deflects to its respective static equilibrium angle by following the typical response of a $2nd$ order dynamic system. Although the simulations start with the airfoil at $0^{\circ}$ pitch angle, where negligible aerodynamic lift or moment is theoretically expected, even a minor asymmetry inherent in the computational grid can induce an initial asymmetric loading over the airfoil. Consequently, the airfoil deflects to a near--zero pitch angle, which subsequently establishes a non--zero lift calculation. Provided that the distance between the pitching axis and the quarter--chord is sufficiently large, and considering that the angular velocity of the resulting response is small, the aerodynamic moment around the pitching axis will overcome the restoring damping moment, causing the airfoil to deflect to progressively larger pitch angles, dynamically heading to the static equilibrium pitch angle. (In order to accelerate this initial phase of the simulations, the initial pitch angle of the airfoils analysed herein is set to $0.2^{\circ}$ nose--up.) As velocity increases to $V=18.4 \; m/s$, the balance between aerodynamic energy input and system damping becomes critical, resulting in marginal stabilisation to the static equilibrium pitch angle. A further increase to $V=18.6 \; m/s$ leads to the observed numerical critical velocity, where self--starting and sustained oscillations are predicted. At this critical velocity, the simulation initially exhibits a bifurcation from the symmetric to the asymmetric aeroelastic mode, subsequently resulting in small amplitude asymmetric oscillations around the static equilibrium angle. For velocities exceeding $V_{c,num}=18.6 \; m/s$, symmetric oscillations of significantly larger amplitude around the zero mean pitch angle are predicted. The phase plane plot presented in Figure \ref{fig:pitch_vs_dpitch} clearly depicts these trajectories, confirming the shift from stable static equilibrium ($V < 18.6 \; m/s$) to stable motion of asymmetric SAO ($V=18.6 \; m/s$) or symmetric LAO ($V > 18.6 \; m/s$).

\begin{figure}
    \centering
    \begin{subfigure}[t]{0.49\linewidth}
        \centering
        \includegraphics[width=\linewidth]{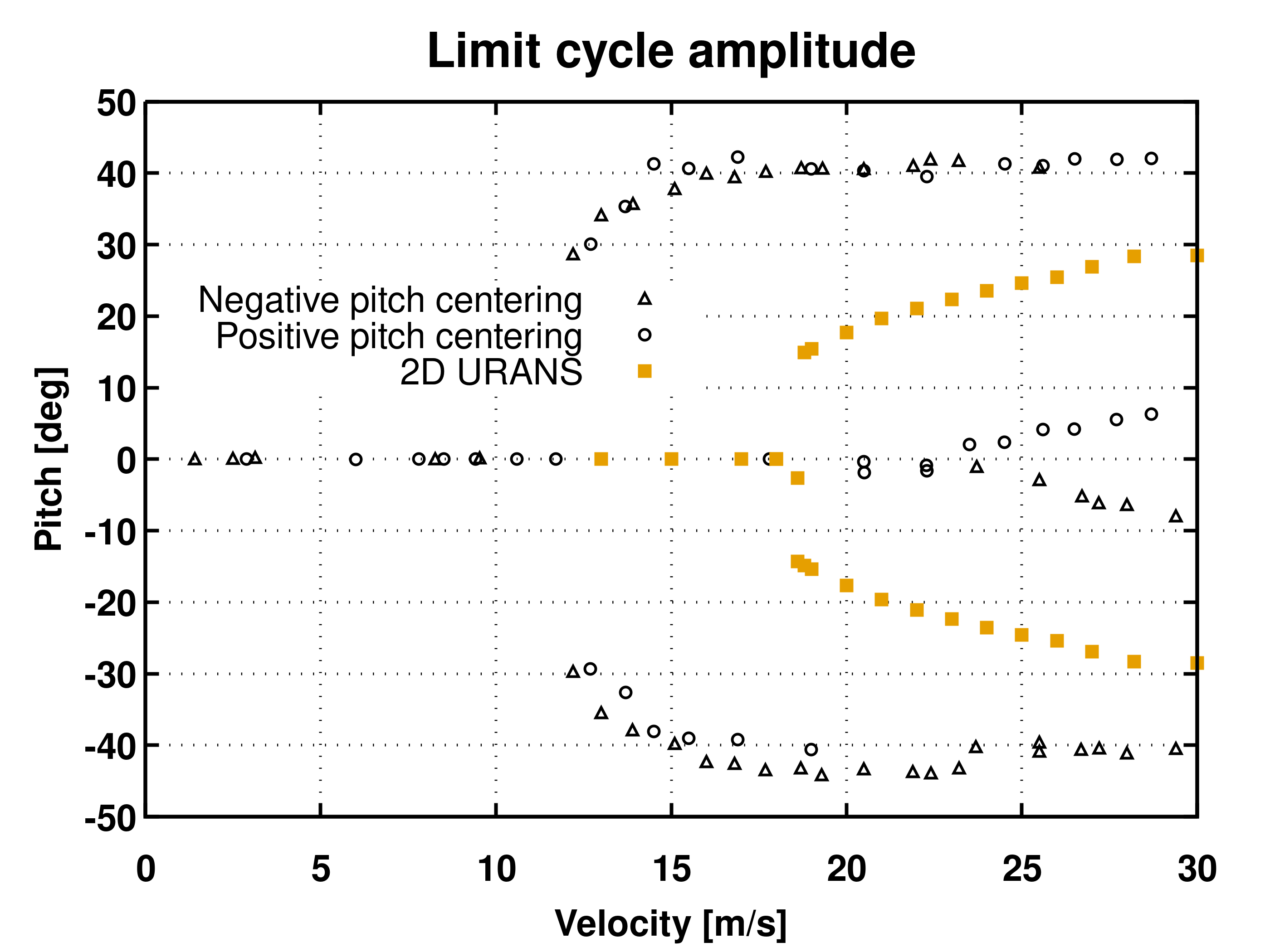}
        \caption{Pitch}
        \label{subfig:pitch_vs_vel}
    \end{subfigure}
    \hfill
    \begin{subfigure}[t]{0.49\linewidth}
        \centering
        \includegraphics[width=\linewidth]{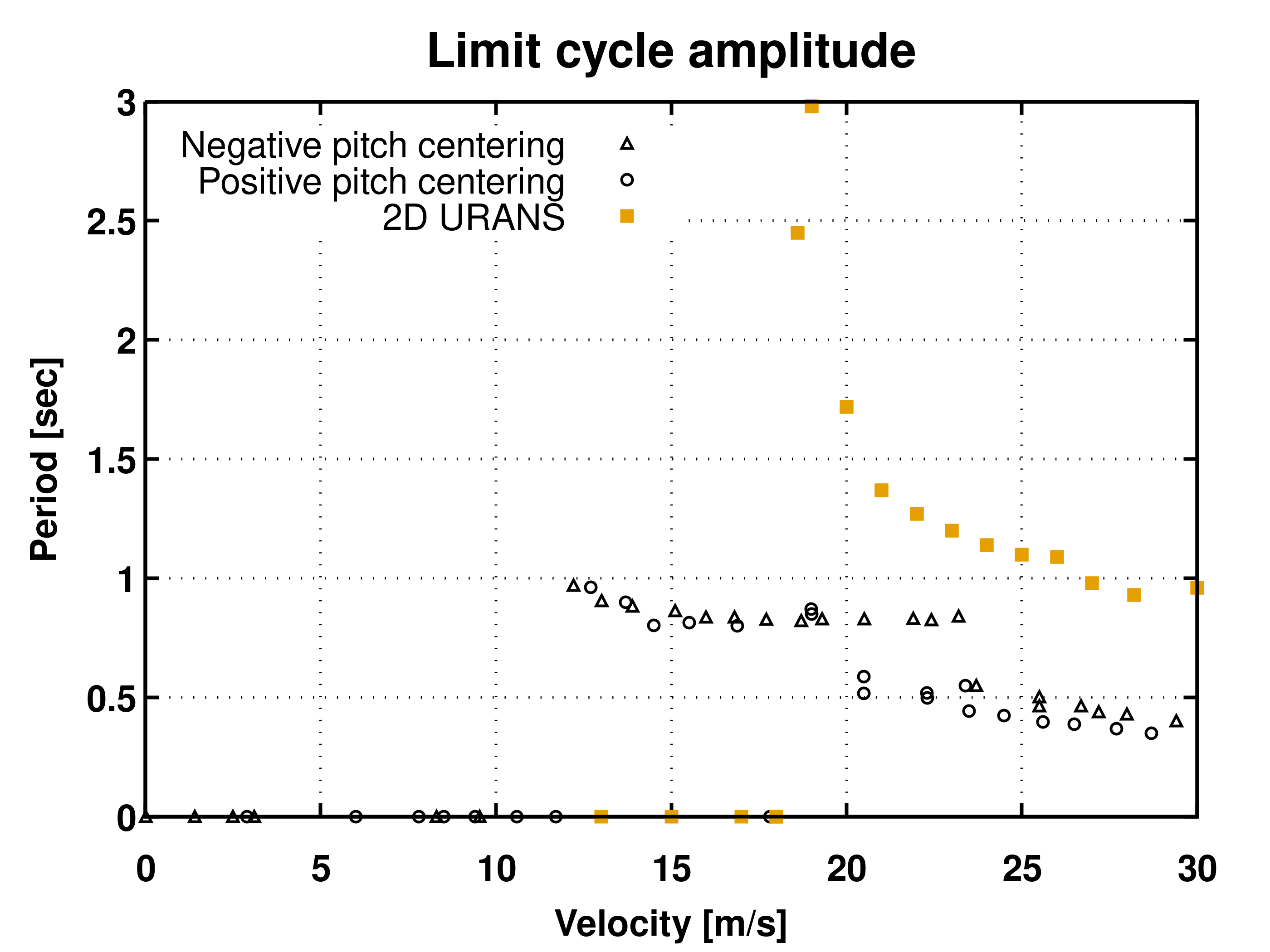}
        \caption{Period}
        \label{subfig:period_vs_vel}
    \end{subfigure}
    \vfill
    \begin{subfigure}[t]{0.49\linewidth}
        \centering
        \includegraphics[width=\linewidth]{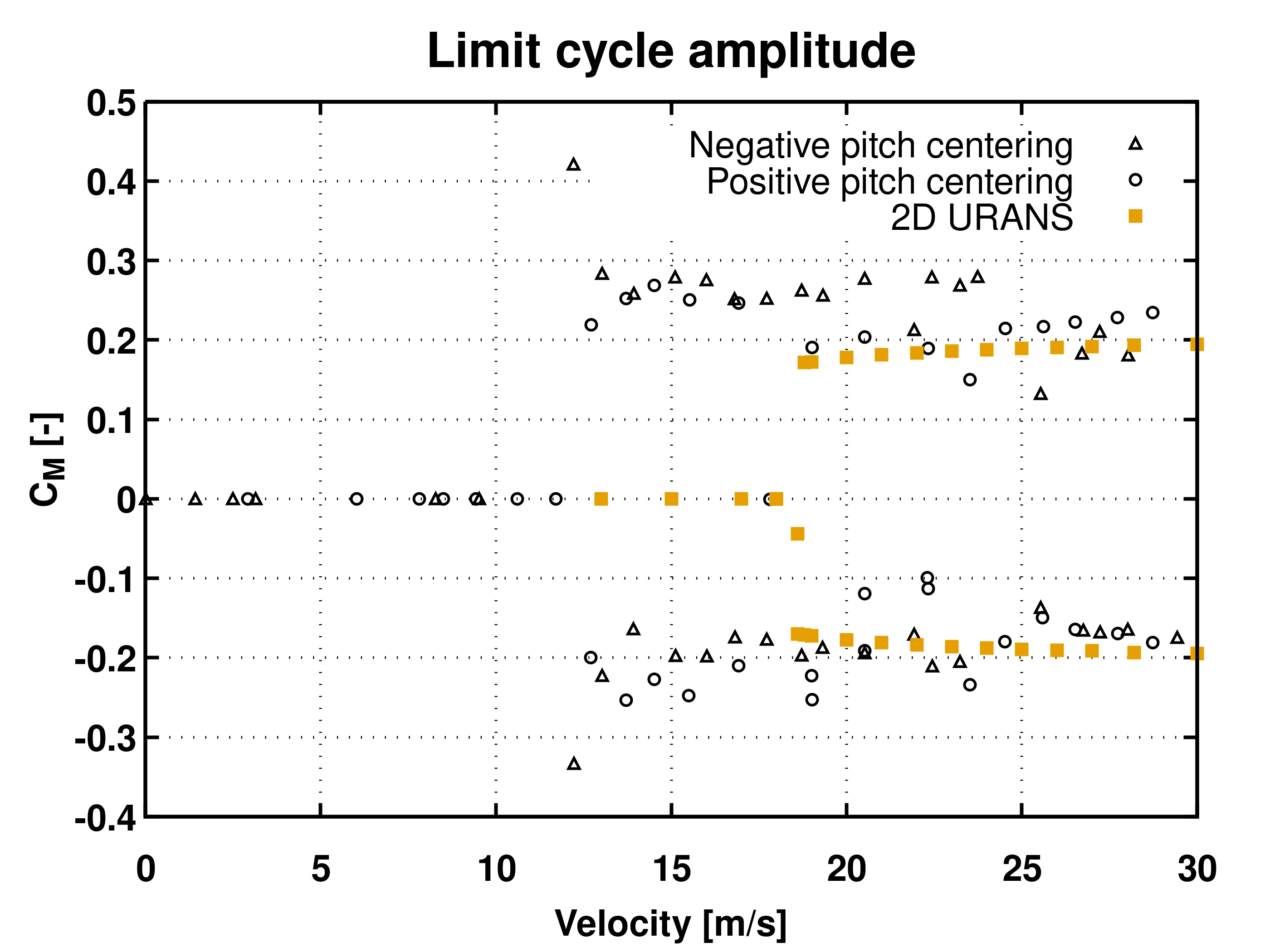}
        \caption{$C_M$}
        \label{subfig:cm_vs_vel}
    \end{subfigure}
    \hfill
    \begin{subfigure}[t]{0.49\linewidth}
        \centering
        \includegraphics[width=\linewidth]{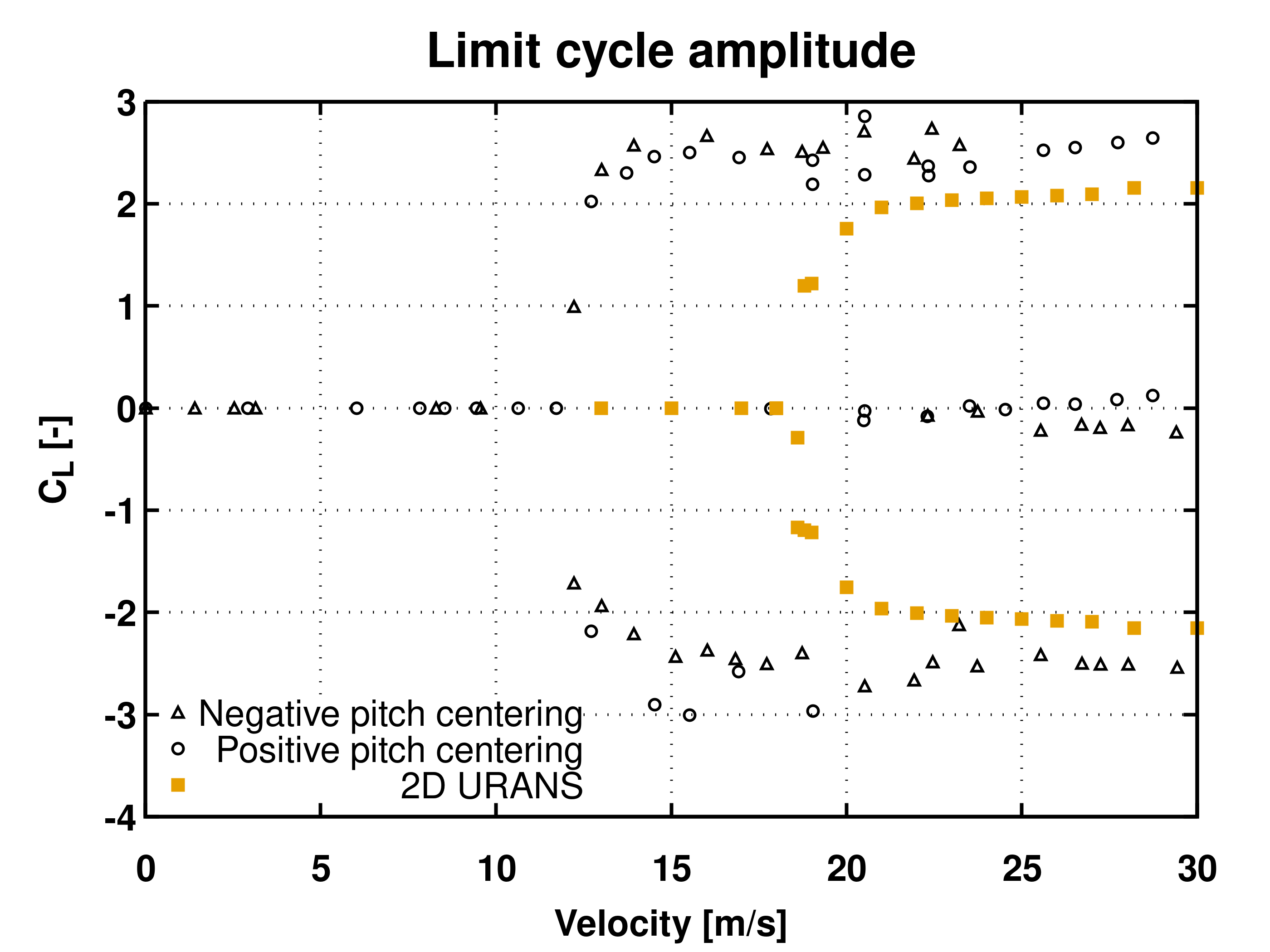}
        \caption{$C_L$}
        \label{subfig:cl_vs_vel}
    \end{subfigure}
    \caption{Comparison of computationally predicted LCO characteristics against experimental measurements from \cite{dimitriadis2009bifurcation}. Figures show the variation with airflow velocity of Pitch Amplitude (a), Period (b), maximum Pitching Moment Coefficient (c) and maximum Lift Coefficient (d). Systematic discrepancies include the over--prediction of the critical velocity ($V_{crit}$) and the under--prediction of the LCO amplitudes across the velocity range, along with the respective larger oscillation Period.}
    \label{fig:velocities_pass}
\end{figure}

Figure \ref{fig:velocities_pass} compares the computationally predicted LCO characteristics (Amplitude and Period variation with respect to the airflow speed) against the experimental measurements. The distinct representation of "Negative pitch centering" and "Positive pitch centering" measured data points originates in a small asymmetry in the experimental wing that drove the zero lift angle of attack to lie between $1^{\circ}$ and $2^{\circ}$, the positive stall angle to be at $13^{\circ}$, the negative stall angle at $10^{\circ}$ and the pitching moment around the quarter--chord to be small, but non--zero at small angles of attack. Hence, in an early phase of the experimental campaign, the wing was centered such that the rest pitch angle was slightly negative, leading to the occurrence of only negative angle asymmetric LCO. The dynamic experiments were then repeated with an alternative centering for the torsional springs that led to a slightly positive equilibrium pitch angle, which in turn led to the observation of positive pitch angle asymmetric LCO. For this reason, only the "Positive pitch centering" symbol is encountered in the positive minimum values of large airflow speeds and only the "Negative pitch centering" symbol in the negative minimum values. Nevertheless, the maximum values agree well with each other on both experimental configurations. On the other hand, the numerical configuration is purely symmetric, yielding in the same absolute value predictions for both positive and negative pitch angles. Yet, for airflow velocities greater than the critical one, only symmetric oscillations around zero mean value are predicted by the numerical model.

\begin{figure}
    \centering
    \begin{subfigure}[t]{0.49\linewidth}
        \centering
        \includegraphics[width=0.9\linewidth]{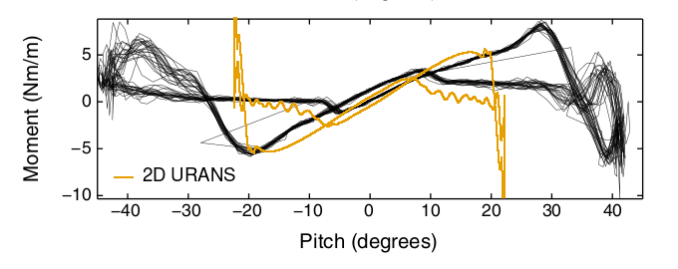}
        \caption{$C_M$}
        \label{subfig:cm_V23ms}
    \end{subfigure}
    \hfill
    \begin{subfigure}[t]{0.49\linewidth}
        \centering
        \includegraphics[width=0.9\linewidth]{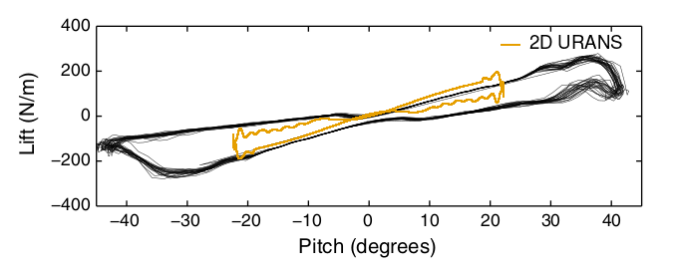}
        \caption{$C_L$}
        \label{subfig:cl_V23ms}
    \end{subfigure}
    \caption{Sectional Pitching Moment Coefficient $C_M$ (a) and Lift Coefficient $C_L$ (b) variation with respect to pitch angle at $V=23 \; m/s$ airspeed. The hysteresis loops demonstrate strong qualitative agreement with wind tunnel measurements, successfully capturing non--linear folding behavior near stall and the extension of the linear phase due to dynamic flow effects.}
    \label{fig:cm_cl_V23ms}
\end{figure}

\begin{figure}
    \centering
    \begin{subfigure}[t]{0.49\linewidth}
        \centering
        \includegraphics[width=0.9\linewidth]{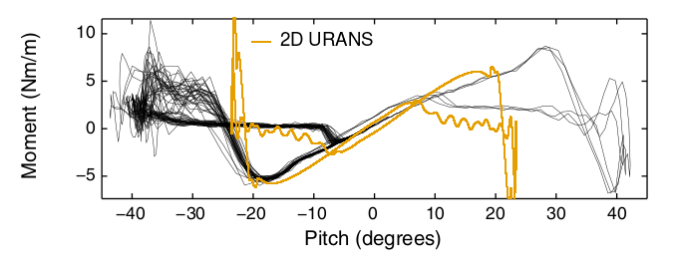}
        \caption{$C_M$}
        \label{subfig:cm_V24ms}
    \end{subfigure}
    \hfill
    \begin{subfigure}[t]{0.49\linewidth}
        \centering
        \includegraphics[width=0.9\linewidth]{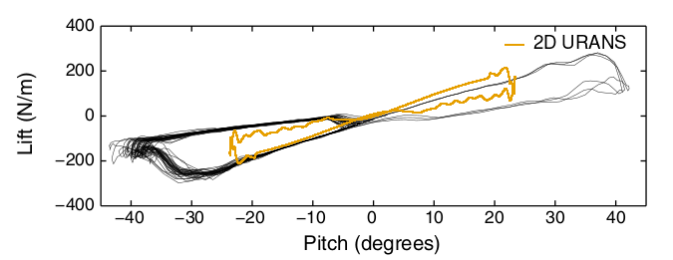}
        \caption{$C_L$}
        \label{subfig:cl_V24ms}
    \end{subfigure}
    \caption{Sectional Pitching Moment Coefficient $C_M$ (a) and Lift Coefficient $C_L$ (b) variation with respect to pitch angle at $V=24 \; m/s$ airspeed. The hysteresis loops demonstrate strong qualitative agreement with wind tunnel measurements, successfully capturing non--linear folding behavior near stall and the extension of the linear phase due to dynamic flow effects.}
    \label{fig:cm_cl_V24ms}
\end{figure}

\begin{figure}
    \centering
    \begin{subfigure}[t]{0.49\linewidth}
        \centering
        \includegraphics[width=0.9\linewidth]{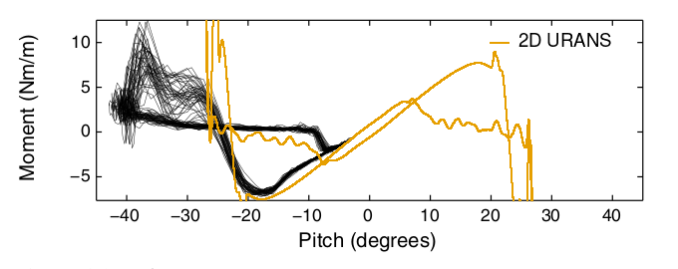}
        \caption{Sectional Moment}
        \label{subfig:cm_V27ms}
    \end{subfigure}
    \hfill
    \begin{subfigure}[t]{0.49\linewidth}
        \centering
        \includegraphics[width=0.9\linewidth]{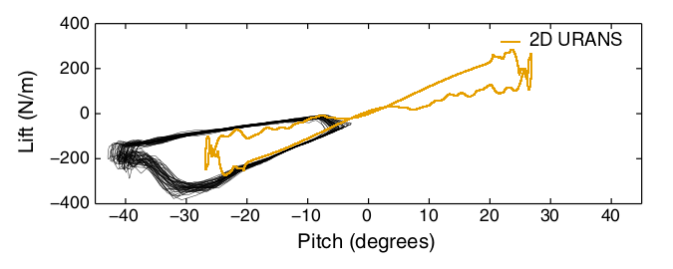}
        \caption{Sectional Lift}
        \label{subfig:cl_V27ms}
    \end{subfigure}
    \caption{Sectional Pitching Moment (a) and Lift (b) variation with respect to pitch angle at $V=27 \; m/s$ airspeed. The hysteresis loops demonstrate strong qualitative agreement with wind tunnel measurements, successfully capturing non--linear folding behavior near stall and the extension of the linear phase due to dynamic flow effects, despite the observation of only the asymmetric part of the loop in the wind tunnel experiment.}
    \label{fig:cm_cl_V27ms}
\end{figure}

\begin{figure}
    \centering
    \includegraphics[width=\linewidth]{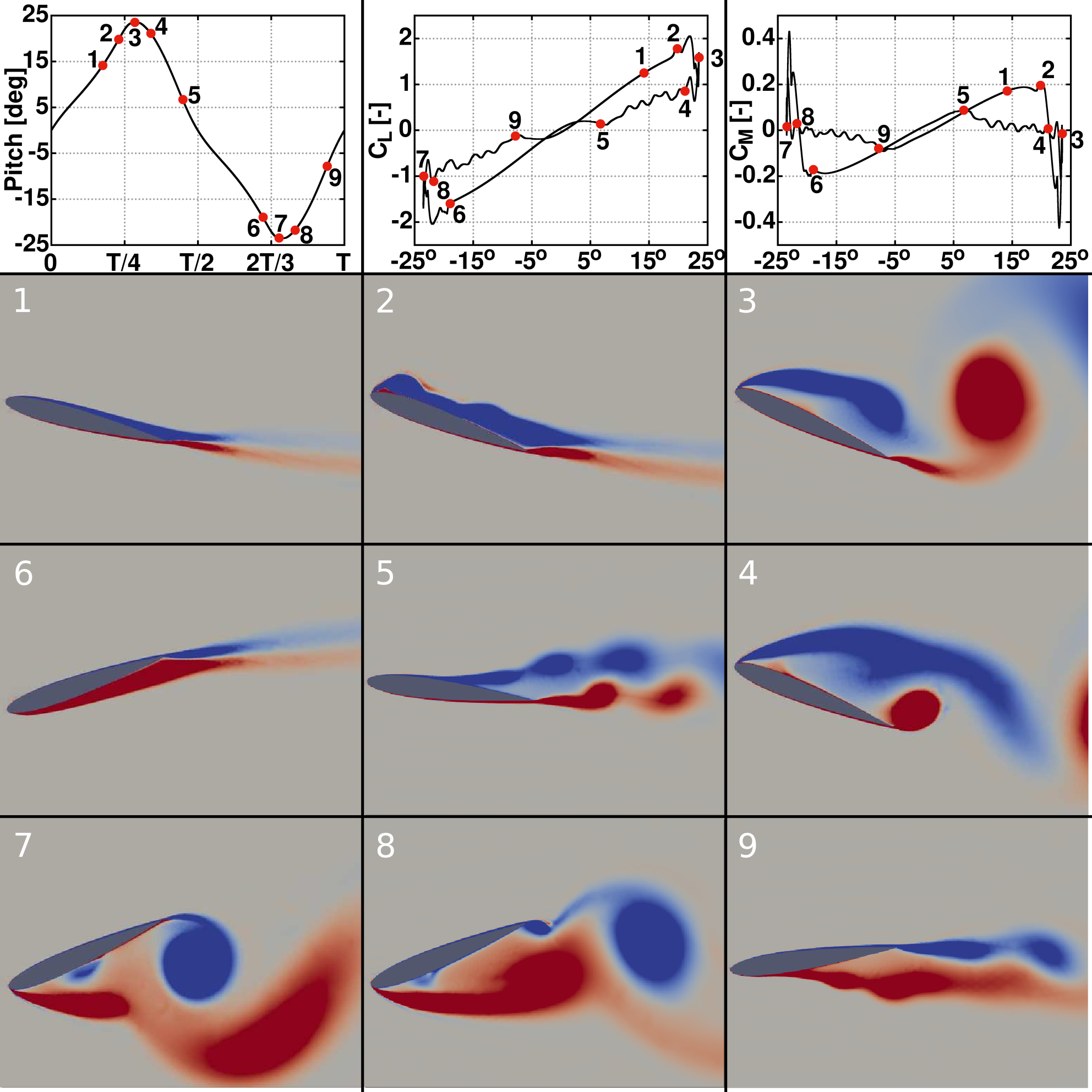}
    \caption{Prediction of sectional Pitching Moment Coefficient $C_M$ and Lift Coefficient $C_L$ variation with respect to pitch angle during one symmetric LCO cycle at $V=24 \; m/s$ (top) and instantaneous vorticity contours (bottom). The sequence illustrates the dynamic stall mechanism, including the generation of a strong LEV during upstroke at positive pitch angles (time instants "$1$" to "$3$"), the shedding and breake--down of the LEV into smaller vorticity structures during the downstroke at positive pitch angles (time instants "$4$" to "$5$") and the respective phenomenon at negative pitch angles (downstroke in time instants "$6$" to "$7$" and upstroke in time instants "$8$" to "$9$"). A strong fluid--structure interaction dynamics is effectively captured by the simulation.}
    \label{fig:flowfield_snapshots_dim}
\end{figure}

Furthermore, the sectional pitching moment and lift hysteresis curves (Figures \ref{fig:cm_cl_V23ms} and \ref{fig:cm_cl_V24ms}) demonstrate strong qualitative agreement with the experimental data. The same outcome can be made also for Figure \ref{fig:cm_cl_V27ms}, where only the asymmetric part of the loop has been measured. Specifically, MaPFlow predicts quite well the minimum values of moment, along with the folding of the curve when the airfoil approaches the minimum pitch value. The disagreement in the respective maximum values of the pitching moment at positive pitch angles originates in the asymmetric shape of the experimental wing. Furthermore, the almost constant values of moment during the upstroke motion at negative pitch angles and the downstroke motion at positive angles is also successfully predicted. The high frequency ripples detected in the predicted moment around these almost constant values are not evident in measurements and are related to the break down of the large LEVs shown in time instants "$4$" and "$8$" to the smaller structures shown in time instants "$5$" and "$9$" of Figure \ref{fig:flowfield_snapshots_dim}, where the dynamic stall mechanism of the flow is illustrated with the help of Vorticity contours. The delay of the boundary layer detachment that causes the extension of the linear part of the aerodynamic loads and the formation of the strong LEV that brings about their sharp drop are fairly resolved by the aeroelastic tool. MaPFlow is also able to predict the extension of the linear phase in post--static--stall angles of attack, by successfully resolving the delay of the boundary layer separation from the airfoil surface during upstroke at positive pitch angles (see time instant "$1$" in Figure \ref{fig:flowfield_snapshots_dim}) and downstroke at negative angles (see time instant "$6$" in Figure \ref{fig:flowfield_snapshots_dim}). The closed loops predicted around $0^{\circ}$ pitch angle (they are more easily identified in moment curves) are also evident in measurements, but slightly shifted due to the experimental wing asymmetry. For the same reason, the measured slope of the linear part of the lift curves at positive pitch angles is slightly lower than that of the negative angles.

In summary, MaPFlow successfully predicts the fundamental dynamics and key features of the aeroelastic system analysed in this Section. Both symmetric pitch oscillations around zero mean value and asymmetric ones around the static equilibrium angle have been successfully predicted. However, the asymmetric oscillations, along with a bifurcation from the symmetric to the asymmetric aeroelastic mode, are encountered by the simulation only close to the critical velocity, where the amplitude of the oscillations is relatively small. The critical velocity is over--predicted compared to the experimentally measured one, which, however,has been found to be a common tendency among various computational tools' predictions also in the SAO Section \ref{ssec:low_amplitude} analysed before. Furthermore, the systematic pitch amplitude under--prediction in all the examined airflow velocities agrees well with the respective larger oscillation period predictions and is attributed to the corresponding lower sectional moment and lift coefficients predicted in the symmetric oscillations velocities (up to $25 \; m/s$). Nevertheless, apart from the narrower pitch angle range, the sectional aerodynamic loads predictions with respect to pitch angle agree qualitatively well with the wind tunnel measurements throughout the computed velocity range.

\section{Conclusion}
\label{sec:conclusions}
The objective of this paper is to perform a comprehensive numerical study investigating the proper prediction of small and large amplitude LCO arising from Stall Flutter instabilities. Computational results are compared against experimental measurements from two distinct campaigns performed at low and moderate Reynolds numbers and numerical predictions by other computational tools of similar fidelity wherever available. The aeroelastic response of the systems was computed using MaPFlow, an in--house compressible URANS solver that employs Low Mach Preconditioning for accurate analysis of low Mach number flow regions, coupled strongly with a rigid body dynamics solver.

The investigation includes a detailed sensitivity analysis of numerical parameters, focusing on grid and temporal discretization, which are critical for accurate FSI predictions. For the grid dependency, a grid  consisting of approximately $105000$ cells ($N_c=1148$ airfoil cells) is selected as the optimal trade--off between accuracy and computational cost. A minimum spacing $\Delta X_{min} = 10^{-4}c$ at the leading and trailing edge has been found to be necessary to achieve a grid--independent prediction for LCO amplitude and period. In the temporal analysis, a non--dimensional time--step value of $\Delta \tau = \dfrac{\Delta t V}{c} = 0.005$, ensuring a minimum of $14400$ steps per period, is required for a time--step independent solution in the aeroelastic context. This is a significantly stricter requirement than that found sufficient for dynamically pitching rigid airfoils, where $1800$ steps per period provide a qualitatively correct numerical prediction.

The computational framework presented herein is able to fairly predict both the aerodynamic loads and the dynamic response of the aeroelastic systems analyzed. Both symmetric pitch oscillations around a zero mean value and asymmetric oscillations around the static equilibrium angle are successfully predicted. Asymmetric oscillations, along with a bifurcation from the symmetric to the asymmetric aeroelastic mode, are observed only near the critical velocity, where the oscillation amplitude is relatively small.

In the case of small amplitude LCO, the employment of the $\gamma$--$Re_\theta$ transition model of Menter is necessary to capture the correct trailing edge separation that leads to LCO, as neglecting it has been reported in literature to result in the complete suppression of the oscillations. MaPFlow, allong with all numerical predictions of the other computational tools presented here, over--predicts the critical velocity for self--starting oscillations. Yet, it manages to fairly capture the pitching motion amplitude and the increasing trend of amplitude and frequency with respect to lower Reynolds numbers. However, the predicted amplitudes are being slightly over--estimated compared to the measured ones in the higher Reynolds number range, yet a good agreement with other computational predictions is shown. This amplitude over--estimation explains the corresponding lower predicted frequencies in the same Reynolds number range, which however tend to successfully increase as the respective amplitudes get reduced.

For LAO, the critical velocity is again over--predicted by MaPFlow. Furthermore, the predicted pitch amplitudes are generally lower than the measured values (predicted plateau at $\approx 29^{\circ}$ versus measured plateau at $\approx 42^{\circ}$), resulting in greater oscillation periods. This under--prediction of amplitude is attributed to the lower aerodynamic loads (excitation) peaks reported in the symmetric oscillations region (up to $25 \; m/s$). Despite these amplitude discrepancies, the sectiotanl pitching moment and lift predictions show qualitatively good agreement with wind tunnel measurements.

In conclusion, the aeroelastic code provides comparable results to those produced by standard CFD solvers of similar accuracy. While the qualitative agreement for complex flow physics like sectional loads and bifurcation behavior is strong.

\section*{Acknowledgments}
This project is carried out within the framework of the National Recovery and Resilience Plan Greece 2.0, funded by the European Union -- NextGenerationEU (H.F.R.I. Project Number: 016749)

The authors gratefully acknowledge the HPC RIVR consortium [(www.hpc-rivr.si)](https://www.hpc-rivr.si) and EuroHPC JU [(eurohpc-ju.europa.eu)](https://eurohpc-ju.europa.eu/) for funding this research by providing computing resources of the HPC system Vega at the Institute of Information Science [(www.izum.si)](https://www.izum.si/en/home/).

The computational grids were generated using the ANSA CAE pre-processor of “BETA\_CAE Systems S.A”.

\clearpage
\bibliographystyle{unsrtnat}
\bibliography{references}  %%% Uncomment this line and comment out the ``thebibliography'' section below to use the external .bib file (using bibtex) .





% \end{thebibliography}

\end{document}